\documentclass[twocolumn]{aastex62}

\newcommand{\beq}{\begin{equation}}
\newcommand{\eeq}{\end{equation}}
\usepackage{amsmath}
\usepackage{longtable}

\shorttitle{Formation of dust clumps and cold shadowed region in gravitationally unstable disk}
\shortauthors{Ohashi et al.}
\graphicspath{{./}{figures/}}

\begin{document}

\title{Formation of dust clumps with sub-Jupiter mass and cold shadowed region in gravitationally unstable disk around Class 0/I protostar in L1527 IRS}

\author[0000-0002-9661-7958]{Satoshi Ohashi}
\affil{RIKEN Cluster for Pioneering Research, 2-1, Hirosawa, Wako-shi, Saitama 351-0198, Japan}
\email{satoshi.ohashi@riken.jp}

\author[0000-0002-1803-0203]{Riouhei Nakatani}
\affil{RIKEN Cluster for Pioneering Research, 2-1, Hirosawa, Wako-shi, Saitama 351-0198, Japan}

\author[0000-0003-2300-2626]{Hauyu Baobab Liu}
\affiliation{Academia Sinica Institute of Astronomy and Astrophysics, 11F of Astro-Math Bldg, 1, Sec. 4, Roosevelt Road, Taipei 10617, Taiwan}

\author[0000-0001-8808-2132]{Hiroshi Kobayashi}
\affil{Department of Physics, Graduate School of Science, Nagoya University, Furo-cho, Chikusa-ku, Nagoya 464-8602, Japan}

\author[0000-0001-7511-0034]{Yichen Zhang}
\affil{RIKEN Cluster for Pioneering Research, 2-1, Hirosawa, Wako-shi, Saitama 351-0198, Japan}

\author[0000-0002-7538-581X]{Tomoyuki Hanawa}
\affil{Center for Frontier Science, Chiba University, 1-33 Yayoi-cho, Inage-ku, Chiba, Chiba 263-8522, Japan}

\author[0000-0002-3297-4497]{Nami Sakai}
\affil{RIKEN Cluster for Pioneering Research, 2-1, Hirosawa, Wako-shi, Saitama 351-0198, Japan}



\begin{abstract}

We have investigated the protostellar disk around a Class 0/I protostar, L1527 IRS, using multi-wavelength observations of the dust continuum emission at $\lambda=0.87$, 2.1, 3.3, and 6.8 mm obtained by the Atacama Large Millimeter/submillimeter Array (ALMA) and the Jansky Very Large Array (VLA).  Our observations achieved a spatial resolution of $3-13$ au and revealed an edge-on disk structure with a size of $\sim80-100$ au.
The emission at 0.87 and 2.1 mm is found to be optically thick within a projected disk radius of $ r_{\rm proj}\lesssim50$ au. The emission at 3.3 and 6.8 mm shows that the power-law index of the dust opacity ($\beta$) is $\beta\sim1.7$ around $ r_{\rm proj}\sim 50$ au, suggesting that grain growth has not yet begun.
The dust temperature ($T_{\rm dust}$) shows a steep decrease with $T_{\rm dust}\propto r_{\rm proj}^{-2}$ outside of the VLA clumps previously identified at $r_{\rm proj}\sim20$ au. Furthermore, the disk is gravitationally unstable at $r_{\rm proj}\sim20$ au, as indicated by a Toomre {\it Q} parameter value of $Q\lesssim1.0$.
These results suggest that the VLA clumps are formed via gravitational instability, which creates a shadow on the outside of the substructure, resulting in the sudden drop in temperature.
The derived dust masses for the VLA clumps are $\gtrsim0.1$ $M_{\rm J}$.
Thus, we suggest that Class 0/I disks can be massive enough to be gravitationally unstable, which might be the origin of gas-giant planets in a 20 au radius. Furthermore, the protostellar disks can be cold due to shadowing.

\end{abstract}



\section{Introduction} \label{sec:intro}

The formation of protostellar disks is an important process not only for the physical processes of accretion to a protostar and removal of angular momentum, but also for the initial conditions of planet formation.
Recent high-resolution observations of molecular lines have revealed the transition of gas motion from accretion to Keplerian rotation toward Class 0/I young stellar objects (YSOs), which allows us to investigate the disk formation process \citep{tob12,mur13,oha14,sak14,yen14,aso15}.
In addition, recent observations using dust continuum emission have revealed the beginning of the formation of structures such as warped, spiral, and ring structures in Class 0/I protostellar disks \citep{she17,she18,sak19,she20,seg20,nak20}.

It has been suggested that ring structures are related to planet formation in the early stage. It is thus important to clarify the formation mechanism.
Various mechanisms for ring formation have been studied, including those that involve unseen planets \citep{gol80,nel00,zhu12}, snowlines of volatile gas species freezing-out onto dust grains \citep{zha15,oku16}, magneto-rotational instability \citep{flo15}, disk winds in unevenly ionized disks \citep{tak18},  a growth front \citep{oha21}, secular gravitational instability \citep{tak14}, and coagulation instability \citep{tom21}.
In addition to the ring formation mechanisms described above, gravitational instability has been theoretically suggested to be important in Class 0/I protostellar disks \citep[e.g.,][]{nak94,vor06,vor10,mac11,tsu17}, and is a possible mechanism for gas-giant planet formation \citep[e.g.,][]{gam01,kra16,vig17}.
The fragmentation of a gravitationally unstable disk and the associated formation of gas-giant planets could be the origin of the existence of planets in Class II protoplanetary disks, as implied by several observations \citep{zha18,tea18,pin18}.
Therefore, understanding the grain growth and gravitational instability in protostellar disks will help us determine when and how planet formation begins.
Note that it has been suggested that the disk masses of Class II protoplanetary disks are insufficient for planet formation \citep[e.g.,][]{man18}, and that planet formation in Class 0/I disks may be important from the perspective of the mass budget problem.

Dust temperature is one of the most important parameters for understanding how and where planets begin to form because it is related to the chemical composition of planetary atmospheres. Dust temperature has been investigated in protoplanetary disks using various molecular lines \citep[e.g.,][]{kam04,not20,obe21,li21,liu21}.
It has recently been suggested that the temperature in Class 0/I disks is hotter than that in Class II protoplanetary disks \citep{van18,zam21,2021ApJ...914...25L}.

To explore the physical conditions for the earliest embedded protostellar system, we study a representative Class 0 low-mass protostar, namely IRAS 04368$+$2557 in L1527 IRS.
L1527 IRS is located in the Taurus molecular cloud at a distance of $d=137$ pc \citep{tor07}, one of the closest star-forming regions. It is a young protostar with a bolometric luminosity of $L_{\rm bol}=2.75$ $L_\odot$ \citep{tob08} and a bolometric temperature of $T_{\rm bol}=44$ K \citep{kri12}. 
In the recent study, they are updated to $L_{\rm bol}=1.6$ $L_\odot$ and $T_{\rm bol}=79$ K, respectively by using the {\it Herschel}-PACS far-IR spectroscopy \citep{kar18}.

The existence of a large Keplerian disk with a size scale of 100 au was reported by \citet{tob12} based on a kinematic analysis of the $^{13}$CO $(J=2-1)$ line.
\citet{oha14} identified a Keplerian disk associated with an infalling envelope based on ALMA observations.
A dramatic change in chemical composition across the centrifugal barrier has also been discovered, which suggests that the disk is still growing \citep{sak14,sak14b}.
\citet{aso17} investigated the rotation and infalling kinematics around L1527 IRS with much higher sensitivity as well as higher angular resolution and estimated the disk radius and stellar mass to be $\sim74$ au and $\sim0.45$ $M_\odot$, respectively.

Recent high-resolution observations have revealed complicated structures of the disk.
\citet{sak19} identified different orbital planes between the inner and outer parts of the disk with a boundary at $40-60$ au, suggesting a warped structure.
In addition, \citet{nak20} identified several clumps inside the 20 au radius of the disk using VLA archival data.
Although the origin of these substructures is still under debate, the VLA clumps may be related to planet formation in the protostellar disk.

Regarding the envelope scale ($\gtrsim100$ au), \citet{oha97} identified an edge-on flattened envelope associated with infalling motion, which was confirmed by \citet{yen13}.
Radio continuum jets have been identified by the VLA observations \citep{loi02,rei04}.
A bipolar outflow was identified in $^{12}$CO molecular lines with single-dish telescopes \citep{tam96,hog98}.
Mid-infrared observations with the {\it Spitzer Space Telescope} and $L'$ band imaging with the Gemini North telescope show bright bipolar scattered light nebulae along the outflow axis on the $10^3-10^4$ au scale \citep{tob08}.
The protostellar disk is highly embedded in the infalling-rotating envelope, proving that the disk around L1527 IRS is still growing.

In this paper, we analyze the multi-wavelength dust continuum emission from sub-millimeter to millimeter wavelengths toward the protostellar disk around L1527 to investigate the origin of the substructures and early planet formation in the disk-forming stage.
We achieved a resolution of $\lesssim0\farcs16$ (corresponding to $\lesssim22$ au) at all observation wavelengths, allowing us to study the detailed structure of the disk.
In particular, the resolution in the disk height direction is $\lesssim5$ au, which is sufficient to discuss the vertical structure of the young growing disk.
The rest of this paper is organized as follows.
The data reduction is described in Section \ref{sec:obs}, and the resulting continuum images are presented in Section \ref{sec:res}. 
In Section \ref{sec:dis}, the physical conditions for the disk, such as inclination (Sec \ref{dis:inc}), low temperature due to the shadowing effect (Sec \ref{dis:temp}), dust scale height (Sec \ref{dis:scale}), grain growth (Sec \ref{dis:grain}), and gravitational instability required for the origin of the substructures, leading to planet formation (Sec \ref{dis:grav}), are discussed.
A summary of the results and a discussion are given in Section \ref{sec:sum}.

\section{Observational Data}\label{sec:obs}

\begin{table*}[ht]
\caption{ List of Observed Data and Total Flux }
\scalebox{0.85}{
\begin{tabular}{lccccccc}
\hline \hline
 Band& Frequency & Total Flux  & Beam size  & rms & rms &robust & Obs ID\\
                     &  (GHz)    &   mJy           &             &  $\mu$Jy beam$^{-1}$ & Kelvin$^1$ &\\
 \hline
Band 7 (Low resolution) & 345 & 430 &0\farcs097$\times$0\farcs086 (P.A. $-21^{\circ}$)  & 87 & 0.11 &0.5 & 2016.A.00011.S$^a$ \\   
Band 4 (Low resolution)& 146 & 56 &0\farcs125$\times$0\farcs040 (P.A. 25$^{\circ}$)  & 8.9 & 0.10 &0.5 & 2019.1.01695.S$^b$\\   
Band 3 (Low resolution)& 92 & 25 &0\farcs158$\times$0\farcs092 (P.A. 0$^{\circ}$)  & 8.5 & 0.085 &0.5 & 2017.1.00509.S$^a$\\   
Q-Band  (Low resolution)& 44 & 3.9 &0\farcs152$\times$0\farcs138 (P.A. 89$^{\circ}$)  & 16& 0.48 &0.5 &  \\   
\hline
Band 7 (High resolution)& 345 & 330 &0\farcs086$\times$0\farcs049 (P.A. 80$^{\circ}$)  & 61 & 0.15 &0.5 & 2019.1.01695.S$^b$ \\   
Band 4 (High resolution)& 146 & 54 &0\farcs097$\times$0\farcs025 (P.A. 21$^{\circ}$)  & 21 & 0.5 &$-2$ & 2019.1.01695.S$^b$ \\   
Band 3 (High resolution)& 92 & 22 &0\farcs089$\times$0\farcs044 (P.A. 2$^{\circ}$)  & 32 &  1.2 &$-2$ & 2016.A.00011.S$^a$ \\   
Q-Band  (High resolution)& 44 & 3.7 &0\farcs050$\times$0\farcs047 (P.A. 80$^{\circ}$)  & 93&  25  &$-2$ \\   
\hline
\label{tabel:flux}
\end{tabular}}
\begin{flushleft}
\tablecomments{
$^1$ The temperature is calculated assuming the Rayleigh-Jeans approximation.\\
$^a$ \citet{nak20}, $^b$ This work
}
\end{flushleft}
\end{table*}

We use the dust continuum data for L1527 taken with ALMA Band 7, Band 4, Band 3, and JVLA Q-Band over the period from 2011 to 2021.
The detailed observations and data reduction for ALMA Band 7, Band 3, and JVLA Q-Band are described in \citet{nak20}.
In this paper, we present new observations of ALMA Band 4 and Band 7 with a higher spatial resolution.
The details are summarized in Table \ref{tabel:flux}.

The proper motion of the target source is large \citep[$\mu_a=0.5$ mas yr$^{-1}$, $\mu_{\delta}=-19.5$ mas yr$^{-1}$;][]{loi02}. 
To allow joint imaging of all data and a comparison of observations, we used the  Common Astronomy Software Applications \citep[CASA;][]{mcm07} task {\it fixplanets} to shift the target source to the expected coordinates on August 1, 2017.

\subsection{ALMA Band 4 Observations}

The L1527 protostellar disk was observed with ALMA Band 4 on September 15 and 21, 2021, with a configuration of C43-9/10.
Four spectral windows with a bandwidth of 1.85 GHz each were used for the dust continuum and molecular lines (e.g., H$_2$CO).
We used all of the spectral windows for the dust continuum emission as a representative frequency for 146 GHz because no molecular lines were detected.
The total bandwidth of the continuum map was 7.4 GHz.
The observations were composed of three tracks, each with an integration time of $\sim40$ min on the source.
J0237+2848 and J0510+1800 were observed as the passband and flux calibrator. J0433+2905 was observed as the phase calibrator.
The antenna configuration covered baseline lengths of 178 to 15238 m.
The maximum recoverable scale ($\theta_{\rm MRS}$) was estimated as $\sim0.6\lambda/L_{\rm  min}$, where $\lambda$ is the observation wavelength and $L_{\rm min}$ is the minimum baseline.
Because the maximum recoverable scale is $\sim1\farcs5$ (corresponding to $\sim200$ au), any structures that extend beyond this size will be resolved out.
The data reduction was performed using a pipeline script provided by ALMA.

\subsection{ALMA Band 7 Observations}

The L1527 protostellar disk was also observed with ALMA Band 7 on September 26, 2021, with a configuration of C43-9/10.
Four spectral windows with a bandwidth of 1.875 GHz each were used for the dust continuum and molecular lines.
We used all of the line free channels for the dust continuum emission as a representative frequency for 345 GHz.
The total bandwidth of the continuum map was 7.5 GHz.
The observations were composed of one track with an integration time of $\sim50$ min on the source.
J0510+1800 was observed as the passband and flux calibrator. J0438+3004 was observed as the phase calibrator.
The antenna configuration covered baseline lengths of 70 to 14361 m.
The maximum recoverable scale ($\theta_{\rm MRS}$) was $\sim1\farcs6$ (corresponding to $\sim220$ au).
The data reduction was performed using a pipeline script provided by ALMA.

We performed self-calibration for the continuum observations in CASA.
The entire imaging process was carried out with the CASA {\it tclean} task.
For constructing CLEAN maps, we adopted Briggs' weighting with a robust parameter of 0.5 and $-2$.
Images with higher sensitivity and a larger beam size were obtained for a robust parameter of 0.5 and images with lower sensitivity and a smaller beam size were obtained for a robust parameter of $-2$.
The beam sizes and sensitivities are summarized in Table \ref{tabel:flux}.

\section{Results} \label{sec:res}

\begin{figure*}[htbp]
\begin{center}
\includegraphics[width=18.cm,bb=0 0 2999 1045]{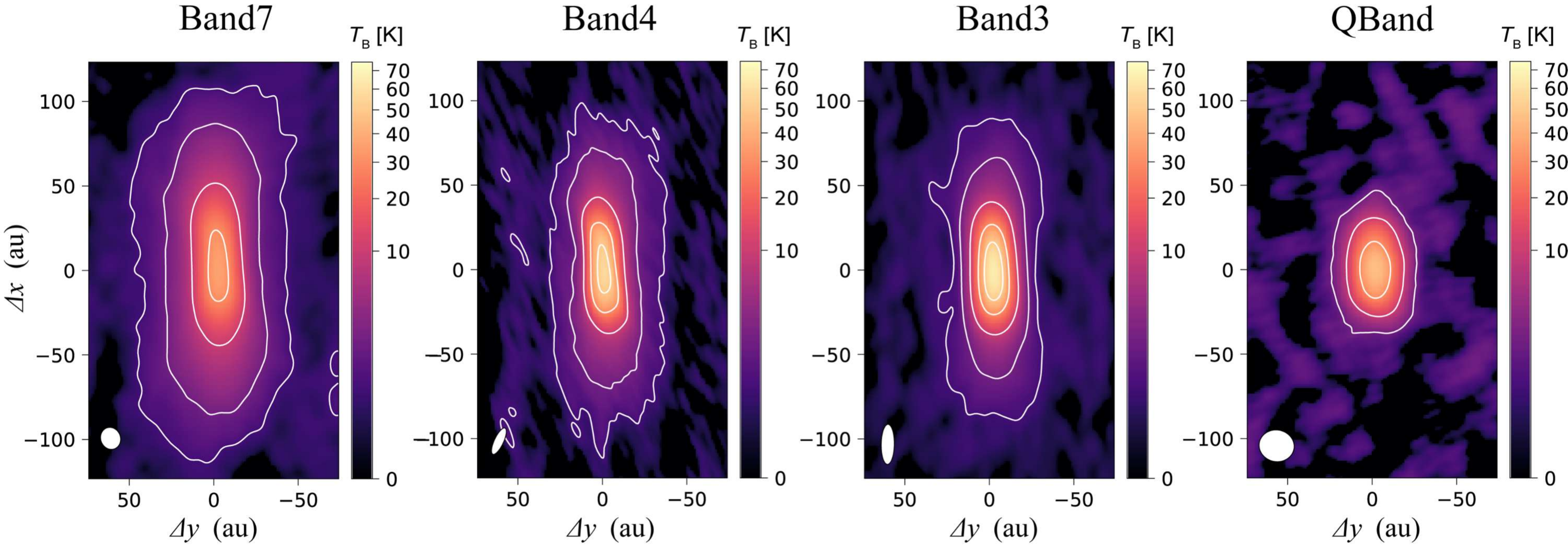}
\end{center}
\caption{Images of protostellar disk around L1527 IRS protostar at ALMA Bands 7, 4, and 3, and VLA Q-Band.  The images were constructed using a robust parameter of $0.5$. The contours are $[0.5, 2, 10, 30]$ K for Band 7, $[0.5, 2, 10, 30, 50]$ K for Band 4, $[0.5, 2, 10, 30, 50]$ K for Band 3, and $[2, 10, 30]$ K for Q-Band. 
In each panel, the resolution is shown by a white ellipse in the lower left.
}
\label{fig1}
\end{figure*}

\subsection{Observed Disk Images}\label{sec:image}

Figure \ref{fig1} shows dust continuum images toward the L1527 disk obtained at sub-millimeter and millimeter wavelengths of 0.87 to 6.8 mm.
The disk structure is well resolved and similar to those reported in previous studies \citep[e.g.,][]{oha14,aso17,nak20}.
We confirmed that the disk is close to being edge-on and is elongated in the north-south direction.
Because the disk has an almost edge-on geometry aligned from north to south, we assume that the $\Delta\delta$ and $\Delta\alpha$ directions correspond to the projected distances in the disk horizontal ($\Delta x$) and vertical ($\Delta y$) directions, respectively.
The projected distance in the disk radial direction ($r_{\rm proj}$) was measured based on the peak position in the Gaussian fit  at each declination ($r_{\rm proj}^2=\Delta x^2 + \Delta y^2$).
The radial direction is mostly consistent with the $\Delta x$ direction.

The dust emission in Band 7 extends to $r_{\rm proj}\gtrsim100$ au, and that in Bands 4 and 3 extends to $r_{\rm proj}\sim80$ au.
In contrast, the Q-Band emission is detected only within $r_{\rm proj}\sim30$ au.
The smaller disk size in the VLA observations is attributed to the sensitivity limit due to weaker dust emission at longer wavelengths.
The extended emission in Band 7 could be from an envelope component as well as disk emission.

The derived peak brightness temperatures ($T_{\rm B}$) for Band 7, 4, and 3, and Q-Band emission are $\sim40$, $\sim60$, $\sim70$, and $\sim50$ K, respectively.
The highest temperature ($\sim70$ K) is found for Band 3 emission, which suggests that Band 7 and 4 emission is obscured in the edge-on geometry. In contrast, the Q-Band emission is affected by beam dilution (see Section \ref{obs:sub}).
The detailed disk structure is discussed in Section \ref{dis:inc}. 
The total flux, beam size, and root-mean-squared (rms) noise level are shown in Table \ref{tabel:flux}.
The total flux for Band 7 emission ($430\pm11$ mJy) is slightly lower than that in previous dust polarization observations ($488\pm14$ mJy) derived by \citet{har18}.  
Because the previous dust polarization observations were conducted with a compact configuration (C40-5), more of the envelope emission may have been recovered compared to that in our observations.

The high spatial resolution and sensitivity of the observations allow the vertical distributions to be determined.
In particular, in the images for Bands 4 and 3, the disk appears to be wider toward the outer edge.
This is consistent with the models developed by \citet{tob13} and \citet{aso17}, who suggested that the disk is flared with $H\propto r^{1.2-1.3}$, where $H$ and $r$ are the dust-scale height and radius of the disk, respectively.
Note that the vertical structure with broadening around $40-60$ au in radius was also reported by \citet{sak17}.

Here, we note that the Band 7 emission is detected over a wide spatial range along the disk vertical direction, extending to $\pm50$ au.
In contrast, the Band 4 and 3 emission is detected within $\pm20$ au.
These different spatial distributions suggest that the Band 7 emission traces not only the disk component but also envelope emission because the Band 7 emission would be sensitive to trace dust thermal emission, and the protostellar disk grows via infalling materials from the envelope.

\begin{figure*}[htbp]
\begin{center}
\includegraphics[width=18.cm,bb=0 0 2999 1285]{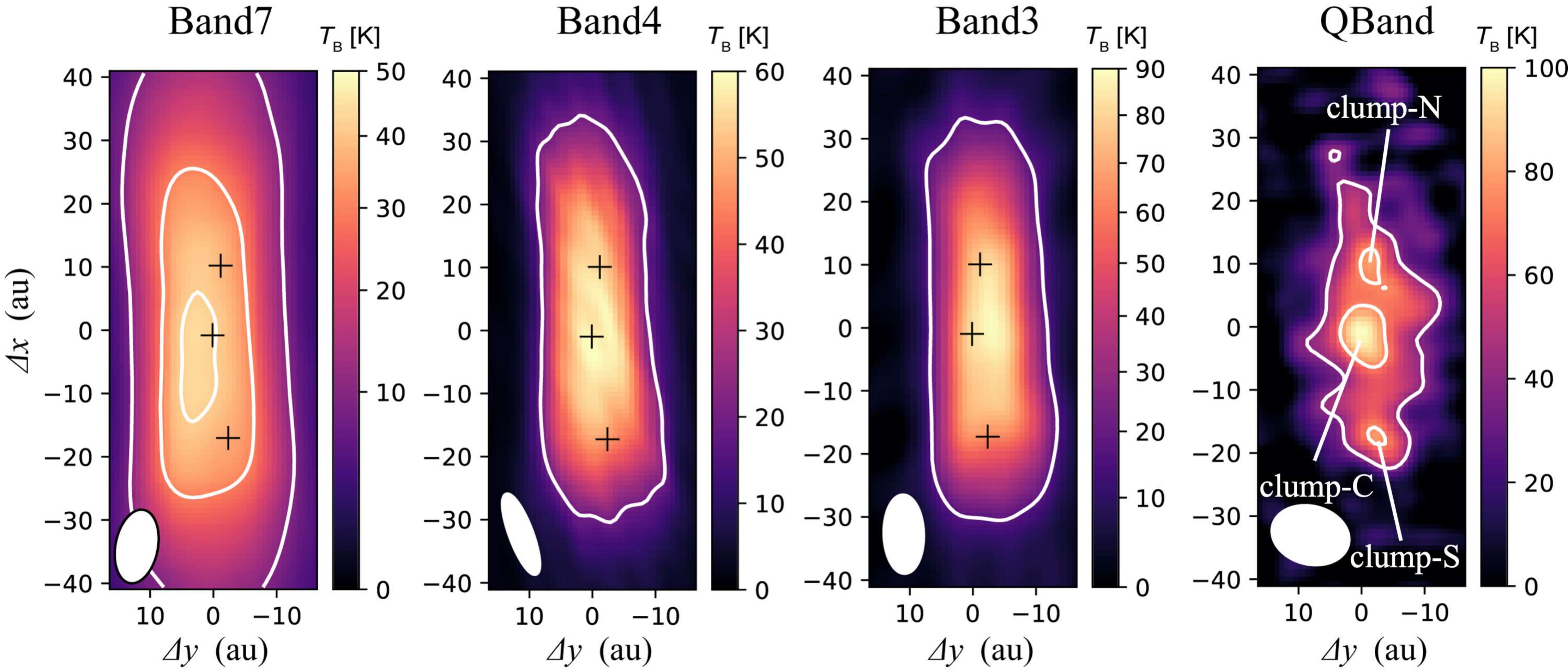}
\end{center}
\caption{Enlarged views of disk around L1527 IRS protostar. The images were constructed using a robust parameter of $-2$, except for the Band 7 image. The ALMA Band 7 image was produced using the extended configuration with a robust parameter of 0.5. The contours are $[10, 30, 42]$ K for Band 7, $[20]$ K for Band 4, $[18]$ K for Band 3, and $[40, 75]$ K for Q-Band. The VLA clump locations (cross symbols) are plotted in the ALMA images.
In each panel, the resolution is shown by a white ellipse in the lower left.
}
\label{fig2}
\end{figure*}

\subsection{Disk Substructures in Inner 40-au Radius} \label{obs:sub}

Figure \ref{fig2} shows enlarged views of the disk images with higher spatial resolutions than those in Figure \ref{fig1}.
The Band 7 image was produced using high-resolution data. The Band 4 and 3 and Q-Band images were produced by applying weighting with a robust parameter of $-2$ in the CASA {\it tclean} task.
We found that about $10\%$ of the total flux was resolved when the weighting was adjusted.
Disk substructures within $r_{\rm proj}\leq40$ au were identified.

The most prominent substructures were identified in the VLA Q-Band observations.
Three clumps (clump-N, -C, and -S) were recognized around $r_{\rm proj}\sim10-20$ au along the disk elongation, which is consistent with \citet{nak20}, who
discussed these substructures as ring or spiral structures in the disk.
We discuss the origin of these substructures in Section \ref{dis:grav}.

The VLA clump locations (cross symbols in the ALMA images) seem to be slightly offset from the disk midplane in the Band 3 image, which may be caused by the correction of the proper motion or absolute position error.
We found that the ALMA images have no counterpart to the VLA clumps even though the spatial resolutions are comparable (the ALMA images have slightly larger beam sizes).
This may have been caused by the high optical depth of the dust emission at the ALMA wavelengths.
The maximum temperature in the VLA observations is $T_{\rm B}\sim100$ K; it decreases to $T_{\rm B}\sim90$ K in Band 3, 60 K in Band 4, and 50 K in Band 7.
The lower maximum brightness temperatures at the disk center for the shorter wavelengths suggest that the shorter-wavelength emission is affected by the higher optical thickness in regions with outer part of the disk, where the temperature is lower. 

\begin{figure}[htbp]
\begin{center}
\includegraphics[width=8.cm,bb=0 0 1364 1592]{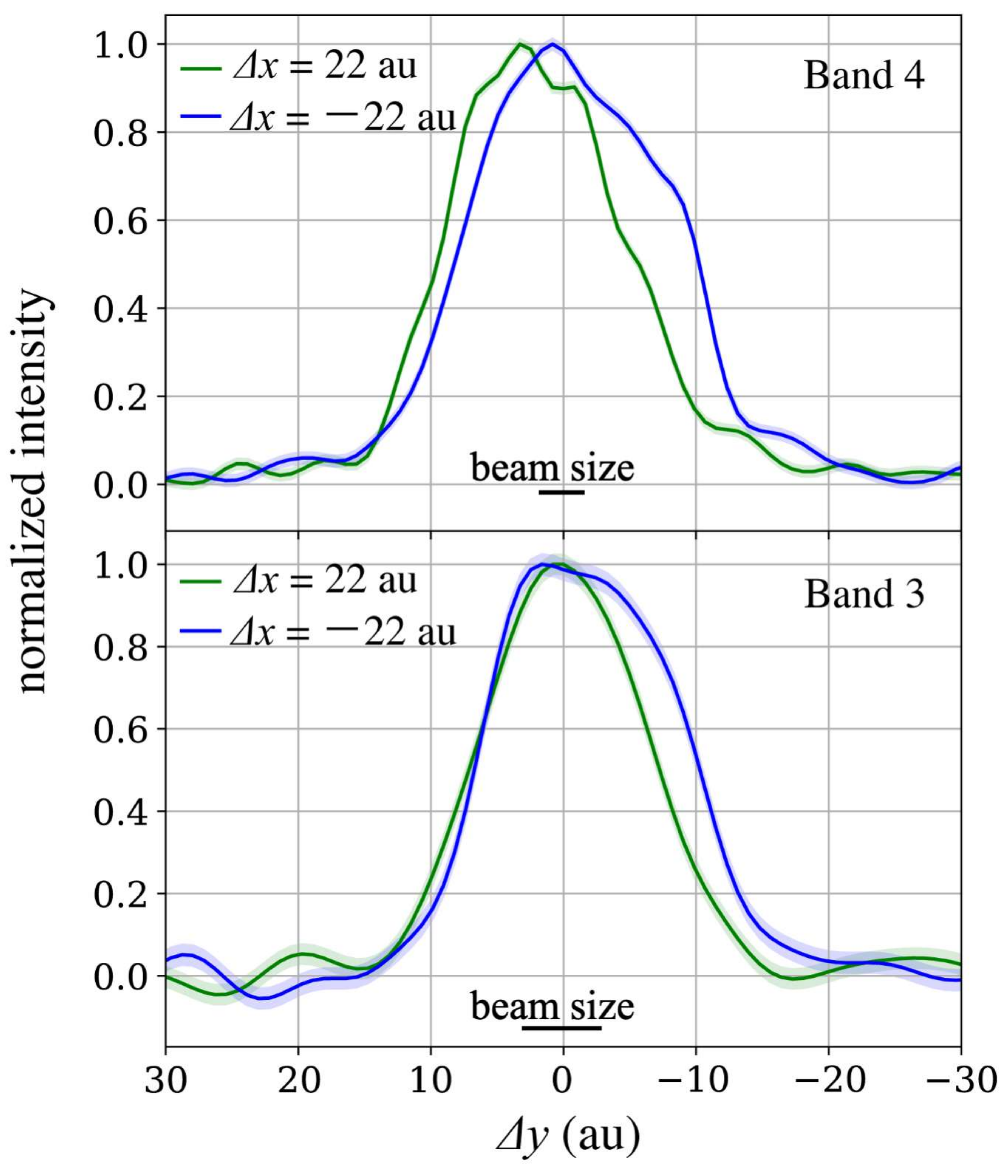}
\end{center}
\caption{Normalized intensity profiles of ALMA Band 4 and 3 emission versus disk vertical direction at $\Delta x=22$ and $-22$ au locations.
The substructures are found in the vertical direction at $\Delta x=-22$ au, where the south VLA clump is located. The shaded regions represent $\pm1\sigma$.
}
\label{fig2dash}
\end{figure}

In addition to the substructures in the radial direction, we resolved the vertical structure of the disk in all bands. 
The Band 7 image shows an asymmetric structure in the brightness temperature even though the VLA clumps are located at almost the disk midplane. The eastern side seems to be slightly warmer than the western side.
We discuss this temperature asymmetry in terms of a flared disk model in Section \ref{dis:inc}.
The Band 4 and 3 images show local enhancements in the vertical direction at the southern VLA clump.
To confirm these substructures, Figure \ref{fig2dash} plots the vertical distribution ($\Delta y$) of the normalized intensity of the Band 4 and 3 images at $\Delta x=-22$ au, where the maximum local enhancements were found, and at $\Delta x=22$ au for comparison.
The Band 4 and 3 intensity distributions both show skewed profiles at $\Delta x=-22$ au.
There is an additional component at $\Delta x\sim-5$ au shifted from the midplane.
The location of  $\Delta x=-22$ au coincide with the southern VLA clump, which suggests that this local enhancement of the disk height is related to the VLA clumps, as discussed in Section \ref{dis:grav}.

\subsection{Spectral Index Analysis}\label{res:alpha}

\begin{figure*}[htbp]
\begin{center}
\includegraphics[width=16.cm,bb=0 0 2999 1375]{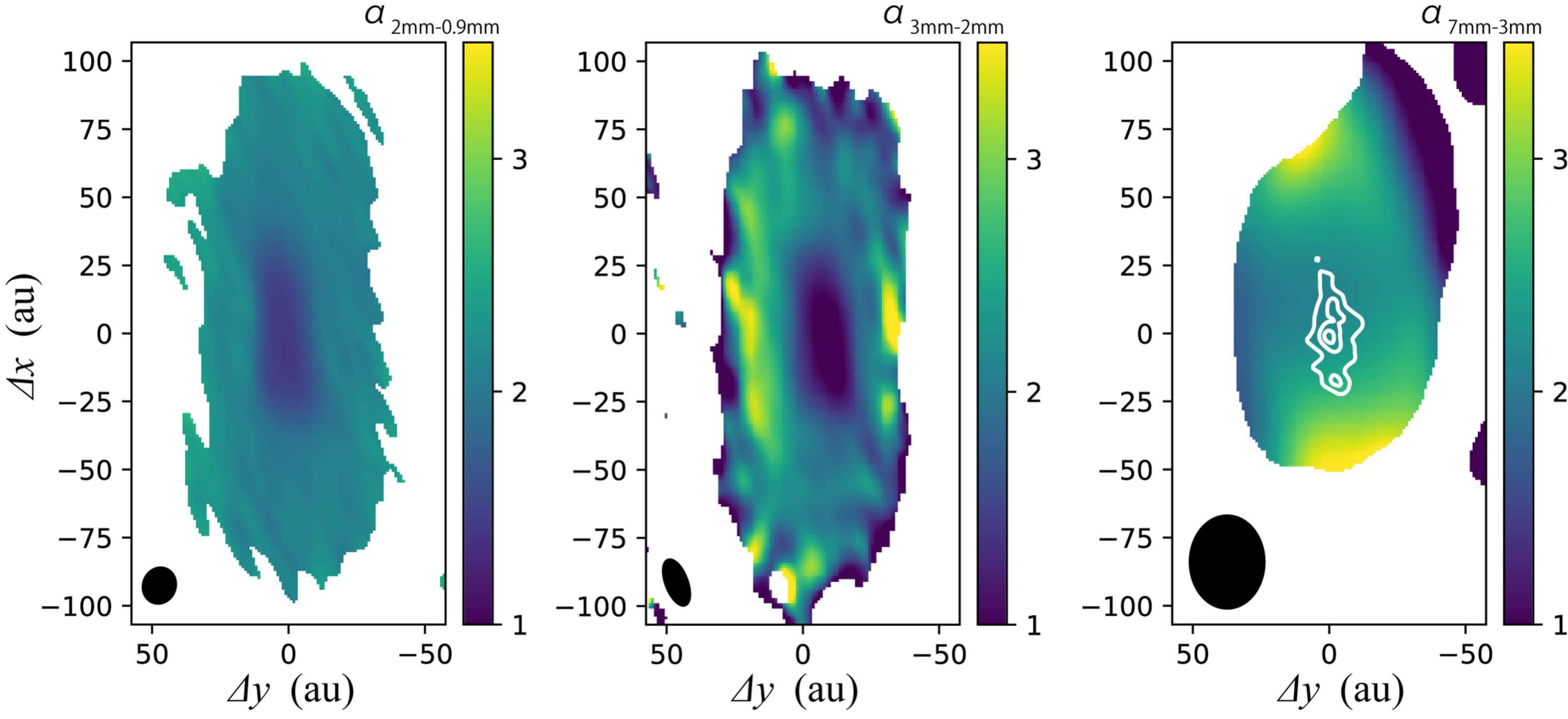}
\end{center}
\caption{Spectral index maps for three combinations derived using multi-frequency synthesis method with $nterms=2$. The spectral indices were derived using two band combinations, namely Band7$-$Band4 ($\alpha_{\rm 2mm-0.9mm}$), Band4$-$Band3 ($\alpha_{\rm 3mm-2mm}$), and Band3$-$Q-Band ($\alpha_{\rm 7mm-3mm}$).
In each panel, the resolution is shown by a black ellipse in the lower left.
}
\label{fig3}
\end{figure*}

The spectral index, $\alpha$, was calculated as
\begin{equation}
\alpha=\frac{\ln I_1 - \ln I_2}{\ln \nu_1 - \ln \nu_2},
\label{eq:alpharadipro}
\end{equation}
where $I$ and $\nu$ are respectively the intensity and frequency at each band. 
The intensity ($I_\nu$) of the dust continuum emission was calculated using the radiative transfer equation.
\begin{equation}
I_\nu=[B_\nu(T)-B_\nu(T_{\rm bg})][1-\exp(-\tau_\nu)],
\label{eq:radipro}
\end{equation}
where $B_\nu$ is the Planck function, $T_{\rm bg}=2.7$ K is the temperature of the cosmic background radiation, and $\tau$ is the optical depth of the dust continuum emission.

We mapped the spectral index $\alpha$ by applying multi-frequency synthesis (mtmfs) in the CASA {\it tclean} task with $nterms=2$.
The three $\alpha$ maps were produced using two band combinations, namely Band 7 $-$ Band 4, Band 4 $-$ Band 3, and Band 3 $-$ Q-Band.
Note that the spectral index for the Band 7 $-$ Band 4 combination may have a first-order Taylor expansion uncertainty because $nterms=2$ may produce a relatively lower spectral index if band separation is wide, as pointed out by \citet{tsu22}. 
Therefore, the spectral index for the Band 4 $-$ Band 3 combination is more reliable.
We note that multi-frequency synthesis for the combination of ALMA Band 3 and VLA Q-Band is not available in CASA.
This may be due to the different data setups for the ALMA and VLA observations.
The $\alpha_{\rm 7mm-3mm}$ map obtained using Band 3 and Q-Band was constructed from the image plane produced with a natural weighting to recover the extended emission.
The total flux of Q-Band increased to 4.8 mJy, indicating that the emission was significantly recovered by natural weighting (see also Table \ref{tabel:flux}).

Figure \ref{fig3} shows spectral index maps for the three combinations.
The $\alpha_{\rm 2mm-0.9mm}$ value is $\lesssim2.0$ in the inner disk ($r_{\rm proj}\lesssim60$ au), which indicates that the emission in Bands 7 and 4 is  highly optically thick.
Even in the outer part of the disk ($r_{\rm proj}\gtrsim60$ au), the $\alpha_{\rm 2mm-0.9mm}$ value is lower than $<2.5$, suggesting that the Band 7 emission is  highly optically thick and the Band 4 emission may be  moderately optically thick.
Although larger dust grains also lead to a lower spectral index, the $\alpha_{\rm 7mm-3mm}$ value being larger than $\alpha_{\rm 2mm-0.9mm}$ and $\alpha_{\rm 3mm-2mm}$ in the entire disk indicates that the low spectral index is due to the optical depth effect rather than dust grain size.
In particular, $\alpha_{\rm 2mm-0.9mm}$ and $\alpha_{\rm 3mm-2mm}$ are as low as $\sim1.5$ in the central region ($r_{\rm proj}\lesssim40$ au).
A spectral index lower than 2.0 suggests that emission at shorter wavelengths traces a colder region.
The optically thick emission allow  only the foreground annulus of the edge-on disk to be observed.
In contrast, emission at longer wavelengths can penetrate the inner radius of the disk because of its lower optical depth, which results in the spectral index being lower than 2.0.
This is confirmed by the different brightness temperatures obtained at different wavelengths in Figure \ref{fig1} and \ref{fig2}.
This effect was previously reported by \citet{gal18}.

The spectral index $\alpha_{\rm 7mm-3mm}$ in the right panel of Figure \ref{fig3} is $\sim2.0$ toward the central region, which may have been caused by the high optical depth. However, $\alpha_{\rm 7mm-3mm}$ becomes as large as 3.6 in the outer radius of $r_{\rm proj}\sim50$ au. 
The $\alpha_{\rm 7mm-3mm}$ value of 3.6 is consistent with that of the interstellar medium (ISMs) \citep[e.g.,][]{dra06}, which suggests a small dust grain size in the outer part of the disk.
The grain size distribution is discussed in Section \ref{dis:grain}.

\subsubsection{Radial Dependence}

\begin{figure}[htbp]
\begin{center}
\includegraphics[width=8.cm,bb=0 0 2389 1605]{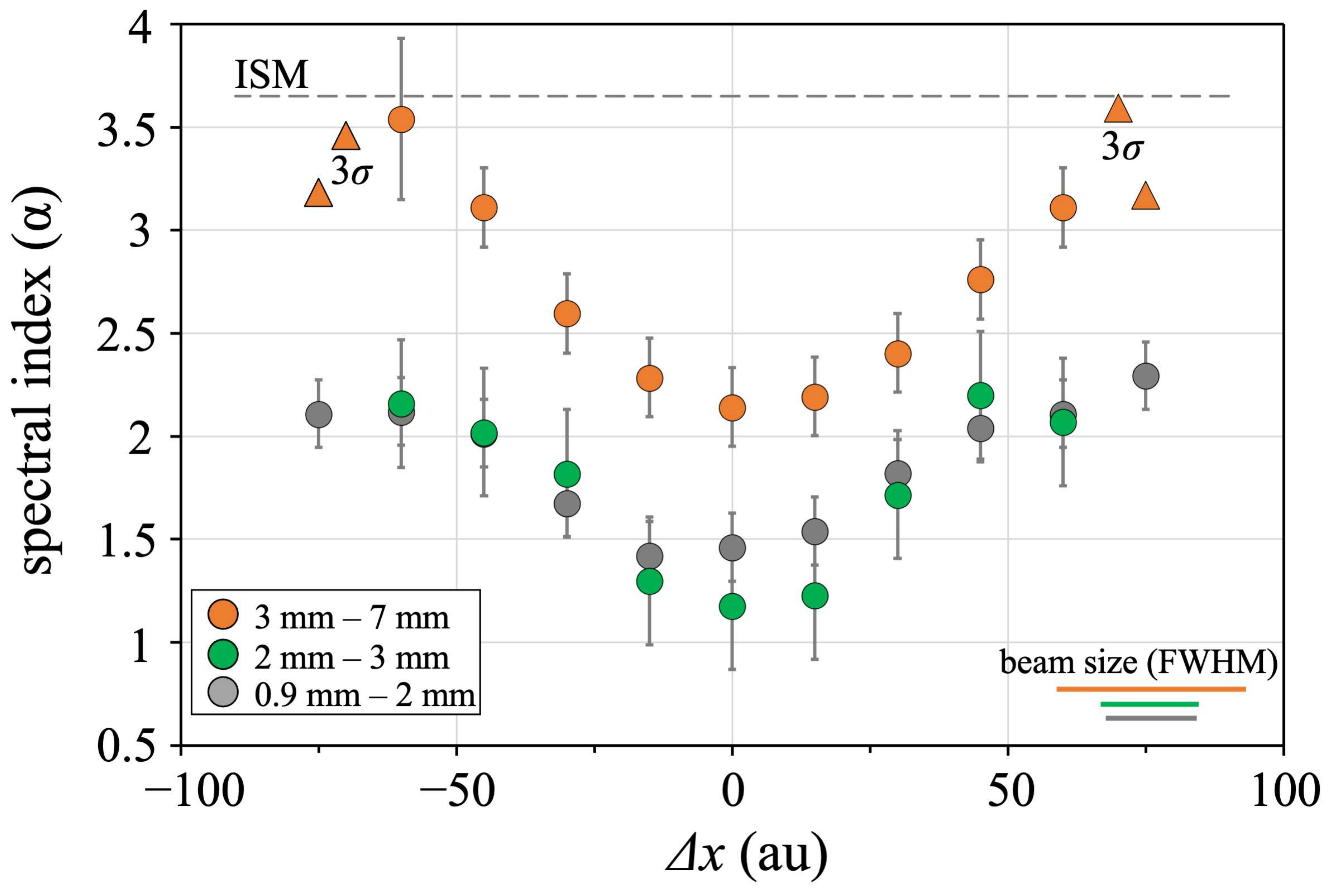}
\end{center}
\caption{Profiles of spectral indices ($\alpha_{\rm 2mm-0.9mm}$, $\alpha_{\rm 3mm-2mm}$, and $\alpha_{\rm 7mm-3mm}$) versus horizontal distance in disk ($\Delta x$) at disk midplane. The error bars represent $\pm1\sigma$, which was calculated using the 5\% flux error and the rms values of the noise levels for the two-band data.
}
\label{alpha}
\end{figure}

To investigate the variation of the spectral index in the disk, we plot the spectral indices as a function of the horizontal distance ($\Delta x$) on the midplane in Figure \ref{alpha}.
We confirmed that $\alpha_{\rm 2mm-0.9mm}$ and $\alpha_{\rm 3mm-2mm}$ are $\lesssim2.0$ in the disk for $\Delta x\lesssim60$ au, indicating that the emission in Bands 7 and 4 is  optically thick.
In contrast, $\alpha_{\rm 7mm-3mm}$ becomes as high as 3.5 at $\Delta x\sim60$ au.
The lower limit of $\alpha_{\rm 7mm-3mm}>3.4$ was derived for $\Delta x\sim70$ au from the non-detection of the VLA Q-Band.
These large spectral index values suggest that the emission in Band 3 and Q-Band becomes  optically thin in the outer part of the disk ($\Delta x\gtrsim60$ au).
Furthermore, $\alpha_{\rm 7mm-3mm}\sim3.6$ is consistent with that for the ISMs, suggesting that grain growth has not yet begun, at least in the outer disk.
We discuss the grain size distribution in more detail in Section \ref{dis:grain}.

\begin{figure}[htbp]
\begin{center}
\includegraphics[width=8.cm,bb=0 0 1413 1083]{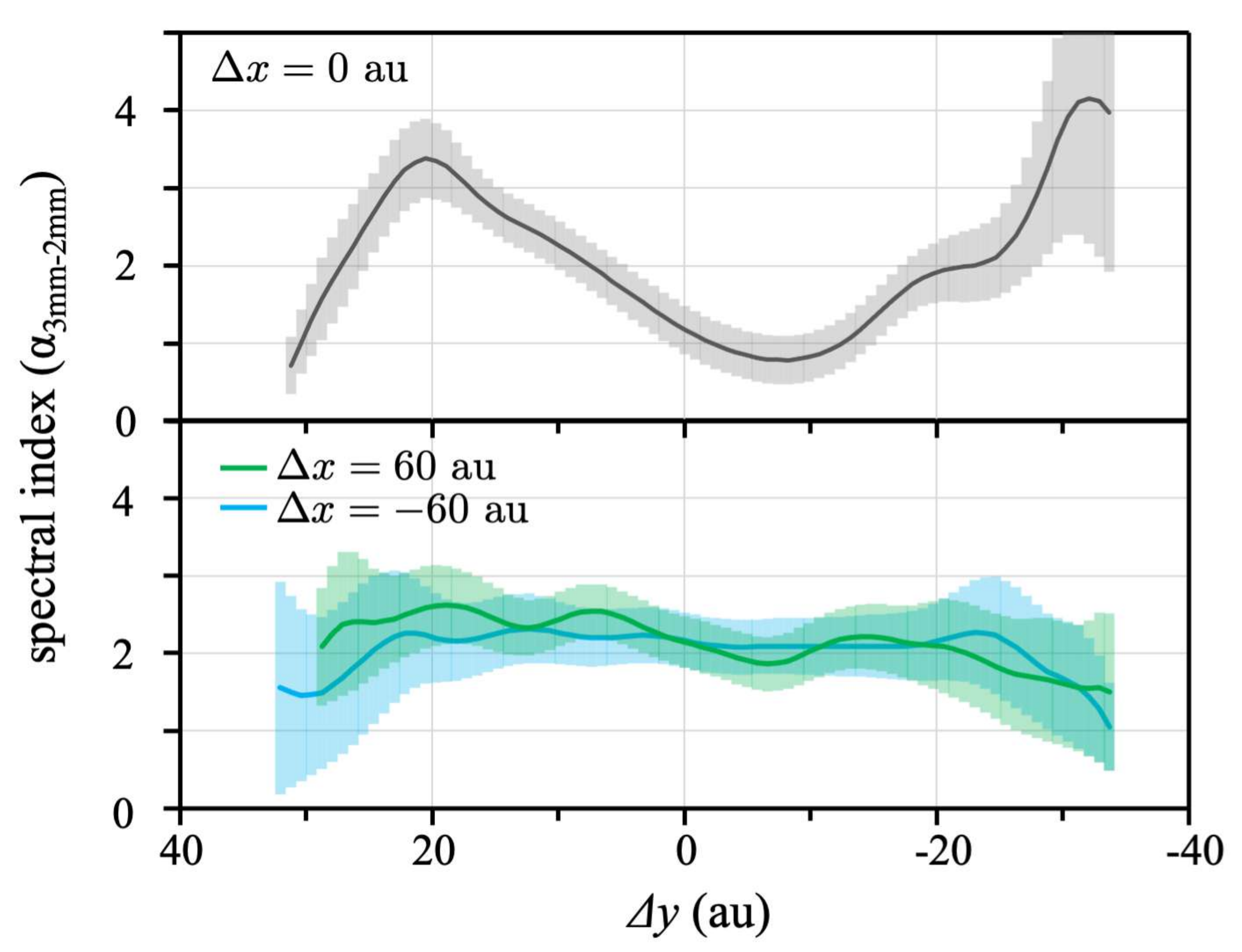}
\end{center}
\caption{Profiles of spectral index ($\alpha_{\rm 3mm-2mm}$) versus vertical distance in disk ($\Delta y$) at $\Delta x=0$ and $\pm60$ au. The shaded regions represent $\pm1\sigma$. The error bars were calculated using the 5\% flux error and the rms values of the noise levels for the two-band data. The data are plotted only where the emission is more than $4\sigma$.
}
\label{alpha2}
\end{figure}

\subsubsection{Vertical Dependence}

Figure \ref{alpha2} shows the vertical distributions ($\Delta y$) of the spectral index $\alpha_{\rm 3mm-2mm}$ at distances of $\Delta x=0$ au and $\pm60$ au.
We found that $\alpha_{\rm 3mm-2mm}$ increases with increasing $\Delta y$ in a radius of $\Delta x=0$ au, which indicates that the emission becomes  optically thinner toward the disk surface. Note that we ignored the inclination effect because the disk has an almost edge-on geometry with an inclination of $\sim5^\circ$.

The peak value of $\alpha_{\rm 3mm-2mm}$ is $\sim3.4$, found at $\Delta y\sim\pm20$ au above the midplane, suggesting that the emission becomes optically thinner and dust grains are not significantly large ($\gtrsim$ mm).
In contrast, $\alpha_{\rm 3mm-2mm}$ shows a constant value of $\sim2.0$ in the vertical direction of $\Delta y\pm30$ au at a radius of $\Delta x=\pm60$ au.
This suggests that the emission remains  optically thick even at $\Delta y\sim\pm20$ au above the midplane.
The outer part of the disk remaining optically thick in the upper layers is consistent with a flared disk structure.
Note that $\alpha_{\rm 3mm-2mm}$ decreases again in $\Delta y\gtrsim20$ au at a radius of $\Delta x=0$ au.
This might be an artifact due to the side-lobe effect. 
The point spread function for our observations show that the side-lobe levels are $\sim0.2$.
Because the peak emission was detected with a signal-to-noise ratio of $>500$, the side lobe may produce artificial emission in the outer region.

The lowest value of $\alpha_{\rm 3mm-2mm}$ was $\sim0.8$, found at $\Delta y\sim-10$ au at a radius of $\Delta x=0$ au.
If the disk is totally edge-on (inclination angle of 0$^\circ$), $\alpha$ is expected to be a minimum at the midplane \citep{gal18}. 
The offset of the lowest potion from the midplane was caused by a slight inclination of the disk.
We discuss disk inclination in the following section (Section \ref{dis:inc}).

\section{Discussion}\label{sec:dis}

We have shown that the dust continuum emission from the protostellar disk of L1527 was detected in a wide range of observation wavelengths (0.87, 2.1, 3.3, and 6.8 mm).
Based on these results, we discuss the disk geometry and physical conditions.
Then, we discuss the origin of the substructures and the early planet formation scenario for a Class 0/I protostar.

\subsection{Disk Inclination}\label{dis:inc}

The protostellar disk of L1527 is nearly edge-on.
The inclination of the disk has been estimated to be $\sim5^\circ$ based on a spectral energy distribution (SED) fitting \citep{tob13} and a kinematic analysis of the infalling-rotating motions of the envelope \citep{oya15}. 
In addition, \citet{oya15} suggested that the western side of the envelope faces the observer, indicating that the western side is the far side and the eastern side is the near side.

\begin{figure}[htbp]
\begin{center}
\includegraphics[width=8.cm,bb=0 0 2296 737]{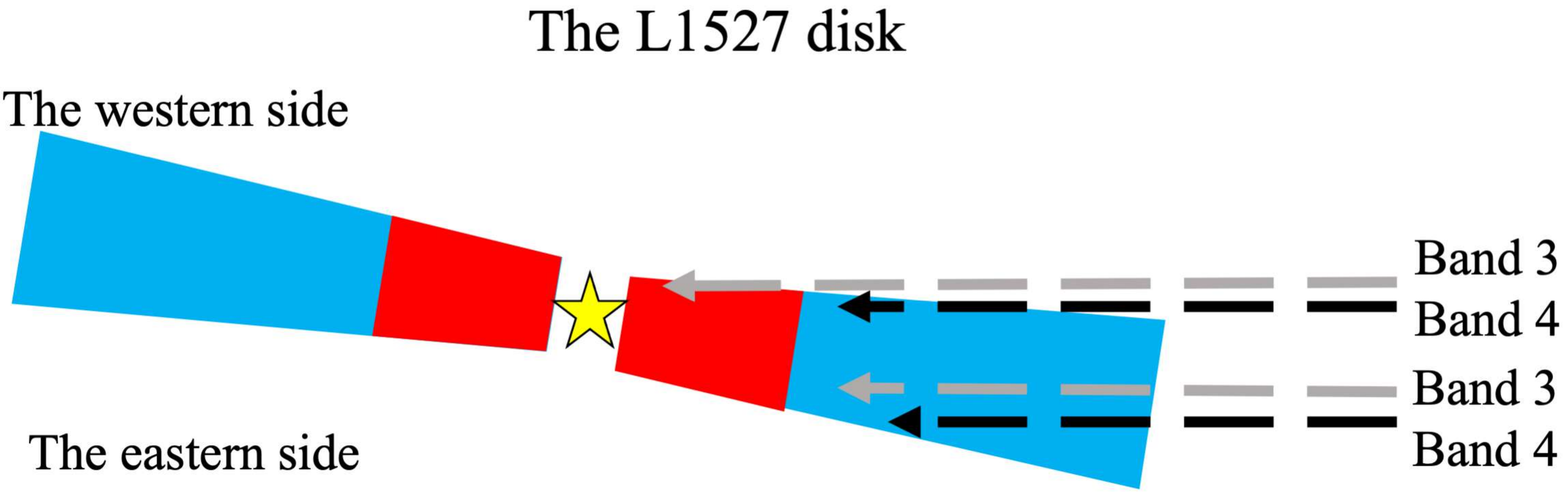}
\end{center}
\caption{Schematic view of inclined disk for explanation of asymmetric temperature profile.
}
\label{view}
\end{figure}

This disk geometry can be assessed using the asymmetric structure of the temperature along the vertical direction of the disk.
If a flared disk is inclined, the near side of the disk will have a temperature gradient where the outer cold region is in front of the inner hotter region.
The far side of the disk will have the opposite temperature gradient.
For an optically thick emission,  the near side should be fainter than the far side because it is obscured (Figure \ref{view}).

\begin{figure}[htbp]
\begin{center}
\includegraphics[width=8.cm,bb=0 0 1628 1614]{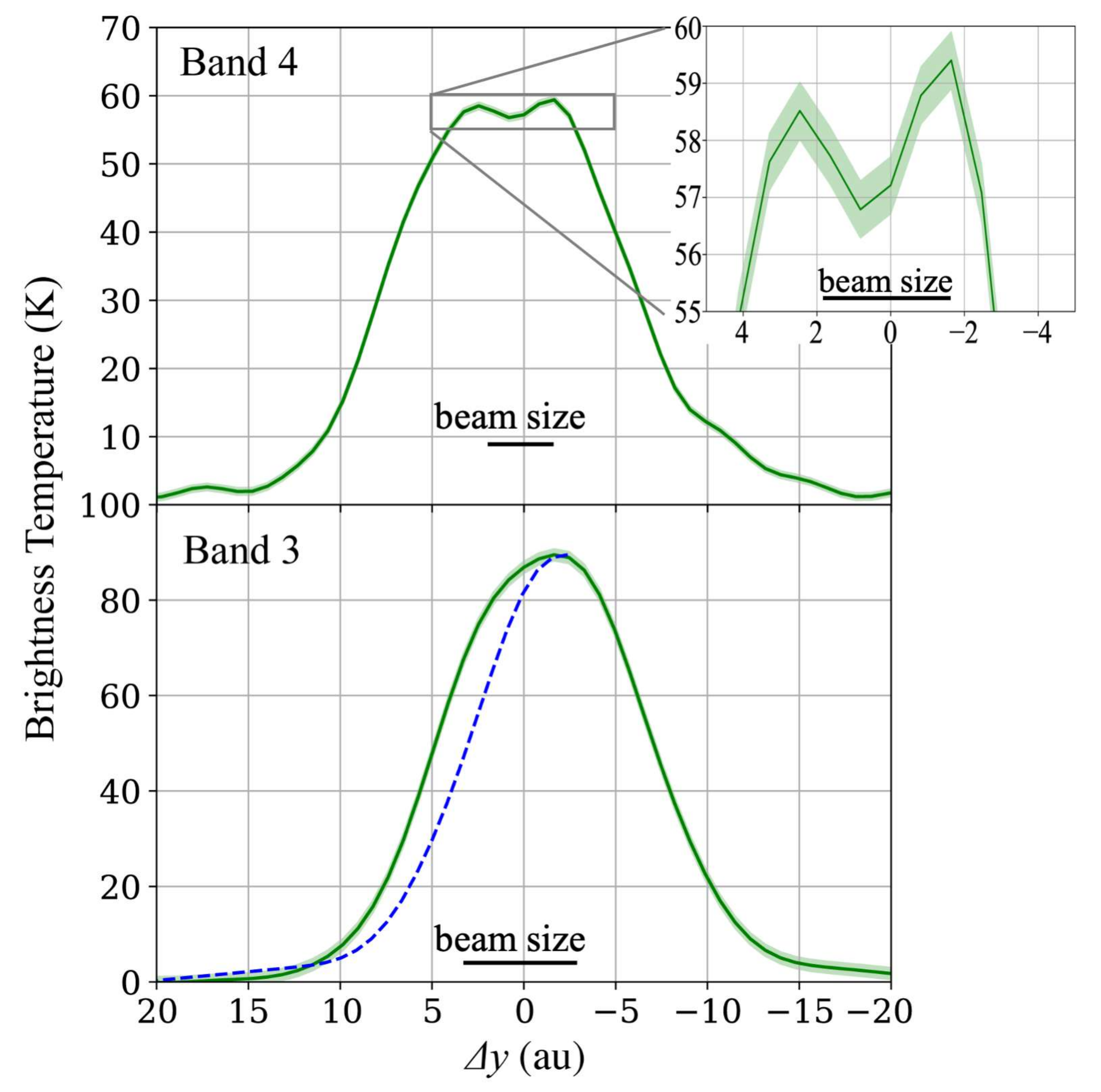}
\end{center}
\caption{Brightness temperatures for Bands 4 and 3 versus disk height  (minor axis). 
The shaded regions represent $\pm1\sigma$.
The blue dashed line shows the profile inverted at the peak temperature to show the asymmetric structure.
}
\label{minor_plot}
\end{figure}

Figure \ref{minor_plot} shows the profiles of the brightness temperatures in Bands 4 and 3 along the disk vertical direction (east-west direction) across the protostar location of $\Delta x=0$ au.
The enlarged profile of the Band 4 emission shows a central dip of $\sim2$ K (corresponding to $\sim4\sigma$). The peak temperature of the western side is slightly higher ($\sim1$ K) than that of the eastern side.
In addition, the Band 3 profile shows that the peak position is slightly shifted to the western side.
These different temperature profiles are caused by the different optical depths of these bands.
Due to higher optical depth toward the central region in Band 4, the emission is efficiently obscured.
Furthermore, the asymmetric profile of the temperature in Band 3 indicates that the Band 3 emission is also optically thick  (but less thick than that for the Band 4 emission).
Similar asymmetric profiles have been found for other Class 0 protostellar disks, such as IRAS 16253$-$2429 \citep{hsi19}, HH 212 \citep{lin21}, and OMC3-MMS6 \citep{2021ApJ...914...25L}.
The temperature of the western side being higher than that of the eastern side indicates that the western side is the far side and the eastern side is the near side of the disk.
The disk geometry suggested by this profile analysis is consistent with that suggested by \citet{oya15}.

The spectral index map could also be affected by the inclination.
As shown in Figures \ref{alpha} and \ref{alpha2}, the lowest position of $\alpha_{\rm 3mm-2mm}$ is shifted by $\sim10$ au to the western side from the midplane.
The Band 3 emission could be highly obscured toward the eastern side (near side) due to the edge-on geometry, and may be emitted from the vicinity of the protostar toward the western side (far side) (Figure \ref{view}).
In this configuration, the western side shows a steep temperature gradient.
Therefore, the lowest value of the spectral index is much lower than 2.0 on the western side.
This scenario should be further assessed using radiative transfer calculations.

\begin{figure}[htbp]
\begin{center}
\includegraphics[width=8.cm,bb=0 0 1622 1069]{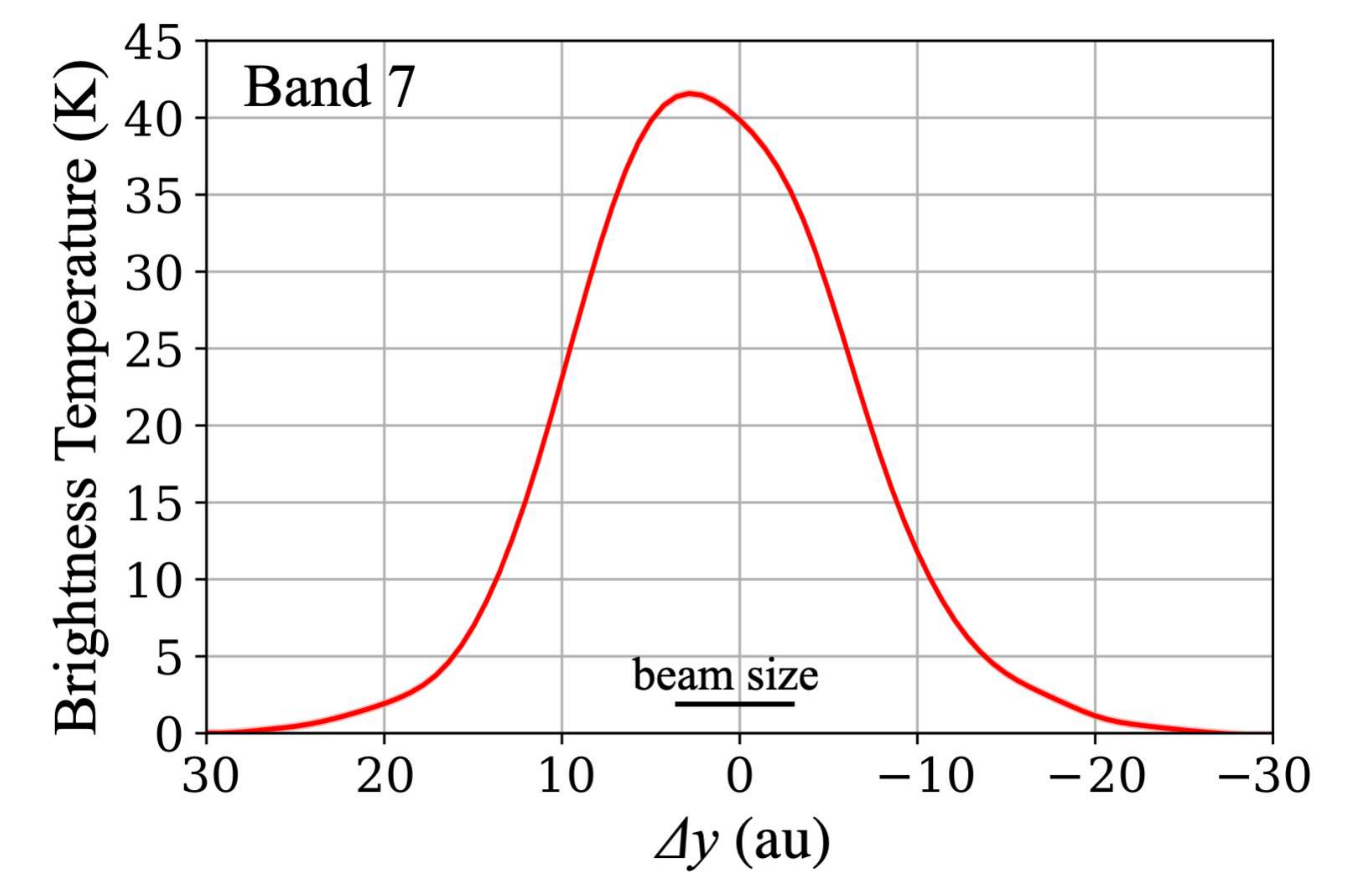}
\end{center}
\caption{Same as Figure \ref{minor_plot} but for Band 7 emission. 
The eastern side of the disk is hotter than the western side, which is opposite to the case in Figure \ref{minor_plot}.
}
\label{b7}
\end{figure}

We plot the temperature profile of the Band 7 emission in Figure \ref{b7} in the same way as that in Figure \ref{minor_plot}.
Interestingly, the temperature profile for Band 7 is opposite to that for Band 3.
The eastern side of the disk is hotter than the wester side for Band 7, which is confirmed by the Band 7 image shown in Figure \ref{fig2}.
If we apply the above discussion to the temperature profile for Band 7, the disk inclination needs to be opposite.
Different inclinations might be the case for this disk because the disk has a warped structure with a boundary at $40-60$ au, as suggested by \citet{sak19}. The warp angle is $\sim5^\circ$.
If the disk is warped, it is also possible that the inclinations of the inner and outer parts of the disk are different .
Because the Band 7 emission is highly optically thick, only the outer part of the disk can be observed, which leads to the different temperature profiles between the Band 7 and Band 3 emission.
However, the Band 7 emission is highly optically thick   and is contributed by the envelope components.
The above discussion may not be applicable to the Band 7 profile.
Further studies of gas kinematics are needed to assess this scenario.

\subsection{Disk Temperature}\label{dis:temp}

\begin{figure*}[htbp]
\begin{center}
\includegraphics[width=18.cm,bb=0 0 2919 1117]{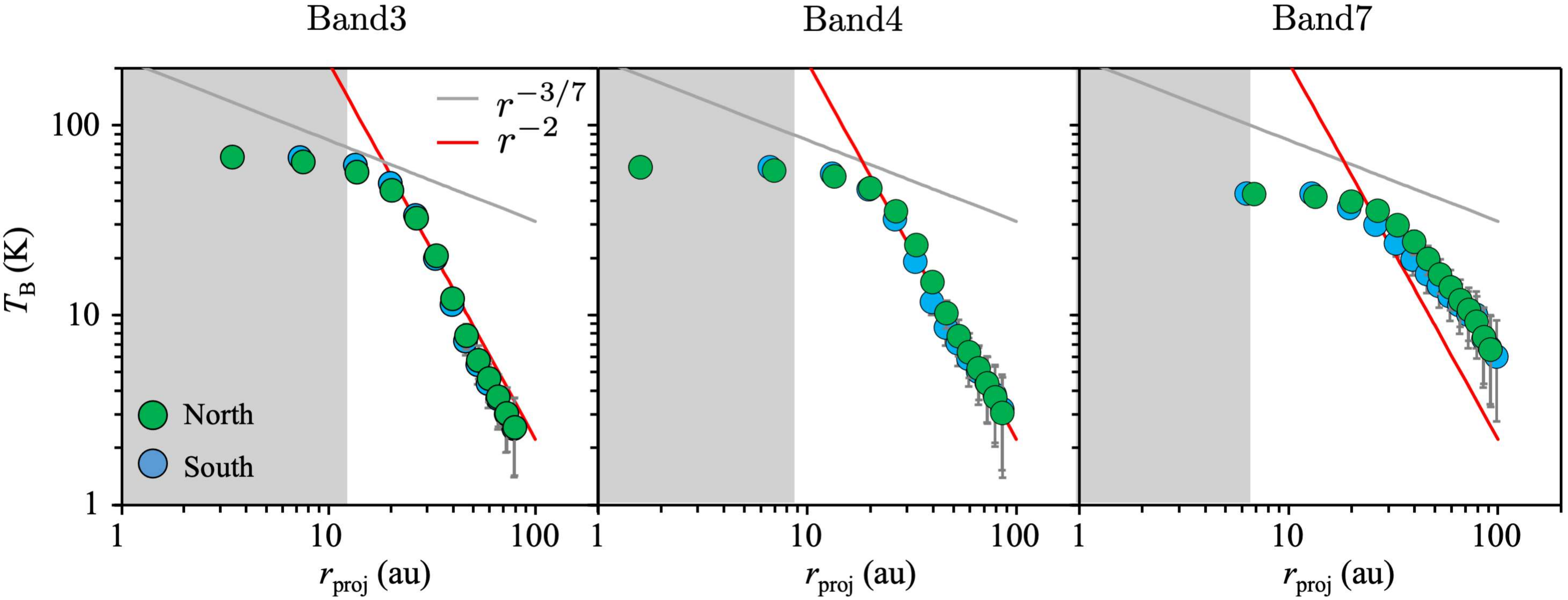}
\end{center}
\caption{Brightness temperatures for Band 3, 4, and 7 emission versus radius based on higher-sensitivity data shown in Figure \ref{fig1}. The gray lines show the irradiation temperature model ($T\propto r_{\rm proj}^{-3/7}$) and the red lines show the power-law best-fit model ($T\propto r_{\rm proj}^{-2}$).
The inner part of the disk is affected by beam dilution, as indicated by the gray regions. The error bars represent $\pm1\sigma$, which was determined using the 5\% flux error and the rms values of the noise levels.
}
\label{temp}
\end{figure*}

\begin{figure}[htbp]
\begin{center}
\includegraphics[width=8.cm,bb=0 0 1110 1675]{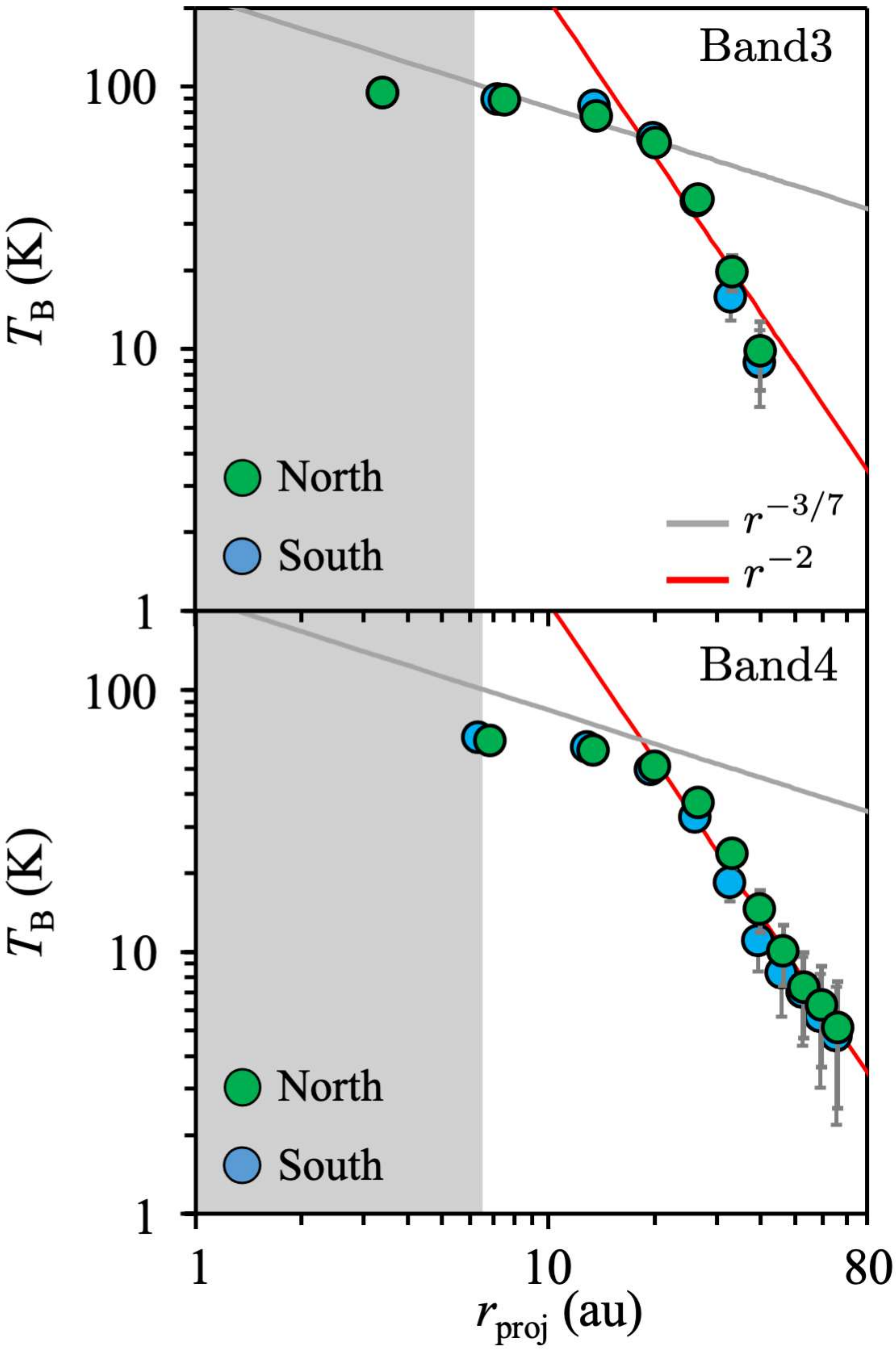}
\end{center}
\caption{Brightness temperatures for Band 3 and 4 emission versus radius based on higher-spatial-resolution data shown in Figure \ref{fig2}. The gray lines show the irradiation temperature model ($T\propto r_{\rm proj}^{-3/7}$) and the red lines show the power-law best-fit model ($T\propto r_{\rm proj}^{-2}$).
The inner part of the disk is affected by beam dilution, as indicated by the gray regions. The error bars represent $\pm1\sigma$, which was determined using the 5\% flux error and the rms values of the noise levels.
}
\label{temp2}
\end{figure}

The multi-wavelength observations of the dust continuum emission allow us to estimate the disk temperature at the midplane.
In particular, the sum of the brightness temperature for an optically thick emission and the cosmic microwave background ($T=2.7$ K) reflects disk temperature (see Equation (\ref{eq:radipro})).
Figure \ref{temp} shows the brightness temperatures for the ALMA Band 3, 4, and 7 emission shown in Figure \ref{fig1} as a function of the projected radius ($r_{\rm proj}$).
The brightness temperature was calculated from the intensity $F_{\rm d}$ (in units of mJy beam$^{-1}$) without the Rayleigh-Jeans approximation.

The brightness temperatures for Bands 3 and 4 are almost the same value for $r_{\rm proj}\lesssim50$ au, indicating optically thick emission. This is confirmed by the low spectral index ($\alpha_{\rm 3mm-2mm}$) in Figure \ref{alpha}.
We also found that the brightness temperature for Band 7 is systematically higher (by a few degrees Kelvin) than that for Bands 3 and 4 in the outer region of the disk ($r_{\rm proj}\gtrsim30$ au).
Taking into account that the Band 7 emission extends beyond than the disk component, as discussed in Section \ref{sec:image}, these results suggest that the Band 7 temperature is contaminated by those of other components.
Envelope emission is a possible contributor because the disk is highly embedded in the parent envelope. If the Band 7 emission consists of the disk and envelope components, the Band 7 temperature is higher than the disk temperature by an amount due to envelope emission.
It may also be possible that the Band 7 emission is contaminated by the emission around the centrifugal barrier, where the temperature is higher due to accretion shock, because the Band 7 emission traces the outer part of the disk.
Unlike in the outer part of the disk, the brightness temperature for Band 3 is higher than those for Bands 4 and 7 in the inner region ($r_{\rm proj}\lesssim20$ au). 
The location where the emission becomes  optically thick is closer to the central protostar at longer wavelengths, resulting in higher brightness temperature for longer-wavelength emission.

The brightness temperature for Band 4 seems to be more reliable for tracing the disk temperature in the region of $30\ {\rm au}\lesssim r_{\rm proj}\lesssim50\ {\rm au}$ compared to that for Band 7 because $\alpha_{\rm 3mm-2mm}\sim2$ in this region and the Band 7 emission are contaminated by the envelope component.
The similar brightness temperatures for Band 3 and Band 4 suggest that the obscuring effect is insignificant.
The sum of the brightness temperature and $T_{\rm bg}=2.7$ K reflects the disk temperature at the midplane because the emission  is optically thick. Note that the beam size in the direction normal to the midplane is sufficiently small to resolve the vertical distribution of the disk.
In contrast, the disk temperature at $r_{\rm proj} \gtrsim50\ {\rm au}$  (outer part of the disk) cannot be determined using the brightness temperature for Band 4 because the emission would become marginally optically thick. 
However, the disk temperature should be $\lesssim10$ K in the outer part because the brightness temperatures for the Band 4 and 7 emission are $T_{\rm B}<10$ K.
On the other hand, \citet{van18} suggested the warm disk scenario in L1527 by the non-detection of the N$_2$D$^+$ ($J=3-2$) emission and the detection of the optically thick emission of the $^{13}$CO ($J=2-1$) and C$^{18}$O ($J=2-1$) lines. The disk temperature is suggested to be above 20 K around a radius of 75 au.
However, the spatial resolutions ($\sim25-130$ au) of these molecular line observations may not be sufficient to resolve the disk. 
In addition, the CO gas would be emitted from the disk surface or envelope.
Therefore, it may be the case that the midplane of the disk is cold ($T\sim10$ K), while the disk surface and the envelope are warm ($T\gtrsim20$ K).

The radial distributions of the brightness temperature show a steep power-law profile in all bands.
The power-law index of $r^{-2.0}$ was found by fitting the brightness temperature for Band 4 in $r_{\rm proj}$ of the $\sim20-90$ au region shown by the red lines in Figure \ref{temp}.
The disk temperature is determined by mainly two heating mechanisms: irradiation and accretion heating. Irradiation from the protostar directly heats the surface and determines the bulk disk temperature \citep{chi97}. Accretion heating is caused by viscous dissipation mediated by turbulence \citep{sha73}.
The disk temperature of these heating mechanisms are suggested to have power-law profiles. In the irradiation heating, the disk temperature shows $T\propto r^{-3/7}$ for optically thick and $T\propto r^{-1/2}$  for optically thin radiation \citep{chi97,chi01}. In the viscous heating, the disk temperature is suggested to be $T\propto r^{-9/10}$  \citep{bau20}.

The slope of $r_{\rm proj}^{-2.0}$ in the L1527 disk is much steeper than those for the irradiation model ($r^{-3/7}$) \citep{chi97,chi01}, optically thin disk model ($r^{-1/2}$), and viscous heating model ($r^{-9/10}$) \citep{bau20}.
The irradiation temperature model is also plotted in Figure \ref{temp}. It was calculated by adopting a luminosity of $2.75$ $L_\odot$ and a stellar mass of $0.45$ $M_\odot$, respectively, as done by \cite{bau20}.
We note that the viscous heating model has a large uncertainty for the temperature profile because of various dependencies, such as the radial mass inflow rate and dust opacity.
One may consider that the steep temperature profiles are caused by the resolved-out effects due to the ALMA extended configurations.
We discuss this effect in Appendix \ref{appen} using simulated observations, based on which we conclude that the steep temperatures are not caused by resolved effects.

The temperature distributions in protoplanetary disks obtained using various molecular lines have recently been reported by the ALMA Large Program MAPS \citep{obe21}.
Using MAPS data, \citet{law21} showed that the temperature at a disk radius of 100 au is $20-30$ K with a radial power-law index of $q=-0.01$ to $-0.23$.
In contrast, the  L1527 disk shows temperatures of $5-10$ K with a radial power-law index of $q=-2.0$.
Although these temperatures were estimated using different observations of the dust continuum emission and molecular lines, the temperature distribution for the L1527 Class 0/I protostellar disk is very different from those for the MAPS protoplanetary disks.
The L1527 disk is colder and has a steeper power-law temperature profile than that for the protoplanetary disks in the outer region ($r_{\rm proj}\gtrsim20$ au) of the disk.

To investigate the temperature distribution within $r_{\rm proj}\lesssim20$ au, we plot the brightness temperatures for Bands 3 and 4 as a function of the radius in Figure \ref{temp2} based on the higher-resolution data obtained with a robust parameter of $-2$.
As shown, the change in the power-law index around $r_{\rm proj}\sim20$ au is not caused by beam dilution.
The temperature profile within $r_{\rm proj}\lesssim20$ au seems to be consistent with the irradiation model for Band 3; the slope of the temperature changes after $r_{\rm proj}\sim20$ au.
Interestingly, the transition radii in the temperature slopes spatially coincide with the locations of the VLA clumps.
Furthermore, the southern part of the outer disk region ($r_{\rm proj}\gtrsim20$ au) is systematically colder than the northern part.
Taking into account that the disk is locally flared in the south VLA clump, the sudden drop in temperature is caused by shadowing of the VLA clumps.

The shadowing effect in protoplanetary disks has been investigated using radiative transfer calculations \citep[e.g.,][]{dul01,dul04,ued19,ohn21,oku22}.
These simulations assumed that a shadow is caused by a planet-induced gap or a dust pile-up at an inner rim due to a dead zone.
They showed that the temperature drops in the shadow region, which is consistent with our result. 
However, for the simulated temperature, the power-law profile returns to normal with $T\propto r^{-0.5}$ in the outer radius.
This is because the outer region is flared and the stellar radiation can heat the flared outer region in the simulations. The absorbed energy at the flared surface is transferred toward the midplane via diffusion.
In contrast, the disk temperature for L1527 maintains a steep slope, $T\propto r_{\rm proj}^{-2}$, to the outer edge of the disk, suggesting that the entire disk may be shadowed by the VLA clumps.

It is worth mentioning that the shadowing of the envelope by the embedded disk was reported in the Class 0/I protostars in the VLA 1623$-$2417 region \citep{mur15}.
They found a cold ring-like structure at the disk-envelope interface in the DCO$^+$ ($J=3-2$) emission and suggested that the temperature drop is caused by the shadowing due to the disk.
These results suggest that the shadowing effect may be a common possible structure in the disks as well as the envelopes.

\subsection{Disk-Scale Height}\label{dis:scale}

\begin{figure}[htbp]
\begin{center}
\includegraphics[width=8.cm,bb=0 0 1614 1465]{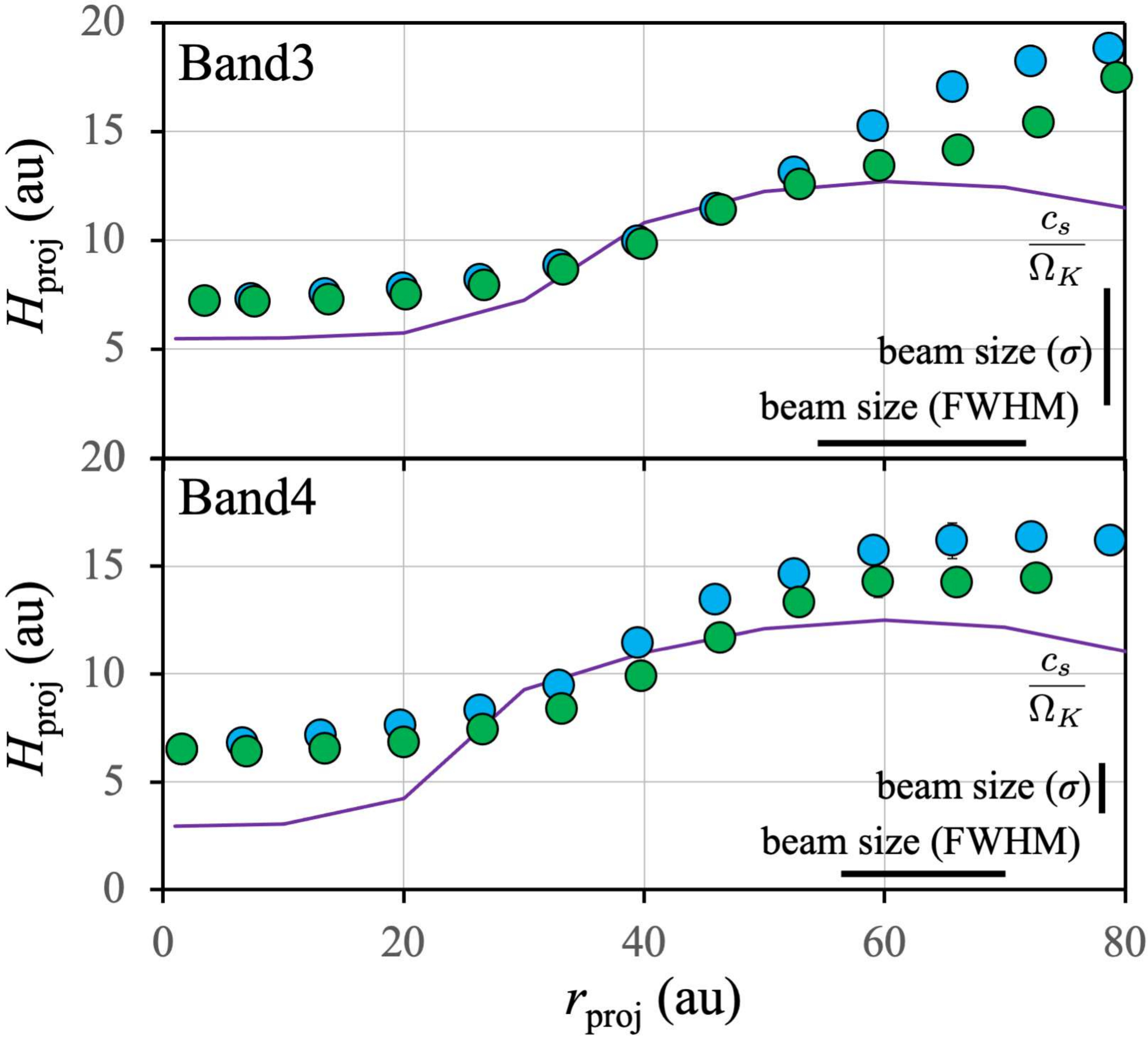}
\end{center}
\caption{Dust-scale height derived from Band 3 and 4 images versus projected radius ($r_{\rm proj}$). The higher-sensitivity images created with a robust parameter of 0.5 were used. The purple lines indicate the gas-scale height estimated using the images from RADMC-3D assuming a power-law best-fit temperature ($T\propto r^{-2}$). The green and blue circles represent the north and south directions, respectively.
}
\label{dust_height}
\end{figure}

The dust-scale height of the disk is an important parameter for constraining the shadowing effect and disk evolution.
The almost edge-on disk geometry of the L1527 disk allows the vertical structure to be investigated.
The projected dust-scale height ($H_{\rm proj}$) was derived from the vertical intensity profile with the standard deviation ($\sigma$) of Gaussian fitting.
Figure \ref{dust_height} shows the projected dust-scale heights derived from the Band 3 and 4 emission for the higher-sensitivity images.
The increases in the dust-scale heights are confirmed in these plots.
The power-law index is roughly estimated to be $H_{\rm proj}\propto r_{\rm proj}^{1.0}$, which is similar to that for previous observations ($r^{1.2-1.3}$) \citep{tob13,aso17}.

It has been suggested that dust grains in protoplanetary disks settle onto the disk midplane due to grain growth and low turbulence \citep{dub95,you07}.
The ratio of the dust- and gas-scale heights allows us to investigate the turbulence strength and grain growth \citep[e.g.,][]{pin16,oha19,vil20,doi21}.
Therefore, we also derive the projected gas-scale heights for comparison. 
The gas-scale height was calculated as 
\begin{equation}
c_{s}/\Omega_{\rm K},
\label{cs}
\end{equation}
\begin{equation}
\Omega_{\rm K} = \sqrt{\frac{GM_\star}{r^3}} = 2.0\times 10^{-7} \Big(\frac{r}{1~\rm au}\Big)^{-3/2}\Big(\frac{M_\star}{M_\odot}\Big)^{1/2}~{\rm s^{-1}},
\label{omega}
\end{equation}
where
 $c_{\rm s}$ is the sound speed, $\Omega_{\rm K}$ is the Keplerian frequency, $G$ is the gravitational constant, and $M_\star$ is the central stellar mass, respectively. 
We assume that the shadowed disk temperature ($T\propto r^{-2}$) is that shown by the red lines in Figure \ref{temp}.
To derive the projected gas-scale height, we performed radiative transfer calculations using RADMC-3D \citep{dul12} assuming an inclination of $5^\circ$ and a position angle of $5^\circ$. Based on the obtained image smoothing with the beam sizes of the Band 3 and 4 observations, we derived the gas-scale heights for the Band 3 and 4 images. The setup was the same as that in \citet{oha19} but the grain size was simply assumed to be $a_{\rm max}=10$ $\mu$m because $\alpha_{\rm 7mm-3mm}$ suggests that the dust grain size is not large. According to the spectral index and grain size relation \citep{bir18}, the grain size is expected to have a range of  $a_{\rm max}\sim0.1-100$ $\mu$m. The emission region was assumed to be optically thin ($\tau<0.1$) regardless of the radius to avoid the obscuring effect of the edge-on disk. 
Note that the gas-scale height is independent of the surface density as long as the emission region is optically thin.

Figure \ref{dust_height} plots the projected gas-scale height (purple solid lines).
We found that the gas-scale height is mostly consistent with the dust-scale height at $r_{\rm proj}\lesssim50$ au, which is consistent with the steep temperature gradient of $T\propto r^{-2}$ and suggests that the dust grains have not settled onto the midplane. No vertical dust settling is consistent with another protostellar disk around the Class 0 protostar HH 212 \citep{lin21}.
The slight differences between the dust- and gas-scale heights in the inner region ($r_{\rm proj}\lesssim25$ au) may be caused by the high optical depth associated with Band 3 and 4 emission. 
The linewidth of the Gaussian fit is larger due to saturation of the emission profile toward the inner radius of the disk midplane.
The substructures of the VLA clumps also contribute to the larger dust-scale heights because additional components are found in the vertical direction, as discussed in Section \ref{obs:sub} and shown in Figures \ref{fig2} and \ref{fig2dash}.

In the outer part of the disk ($r_{\rm proj}\gtrsim60$ au), the dust-scale height seems to be larger than the gas-scale height.
\citet{sak17} found a similar trend, with the CCH emission showing a flared disk structure.
Their interpretation of this result is that the accretion gas stagnates around the centrifugal barrier and moves in vertical directions.
This scenario is possible for the dust because the gas and dust motions are coupled in the disk-forming region.

The increase in dust-scale height may also be caused by inclination, temperature, and outflow contamination.
For example, if the disk is slightly face-on at $r\geq60$ au, the projected scale height becomes larger. This scenario might be consistent with the warped structure because the transition of the inner and outer orbits has been suggested to be $r_{\rm proj}\sim40-60$ au \citep{sak19}.
It is reasonable that the warped structure may change the disk inclination.
The temperature distribution may also increase the scale height of the gas and dust because the accretion shock caused by infalling materials from the envelope may raise the temperature around the centrifugal barrier \citep{sak14}.
The outflow emission may also affect the dust distribution because energetic outflows have been observed toward L1527 \citep{tam96,hog98,tob08,tob10,flo21}.
These outflows are launched perpendicular to the disk elongation.
Recent three-dimensional magnetohydrodynamical simulations show that dust grains can also be launched by outflows, similar to gas motion \citep{tsu21}.
Therefore, it is possible that the dust-scale height is affected by the entrainment of outflows.
To investigate these scenarios, detailed modeling and molecular line observations are needed to reveal the gas kinematics in the disk.

\begin{figure}[htbp]
\begin{center}
\includegraphics[width=8.cm,bb=0 0 1713 1681]{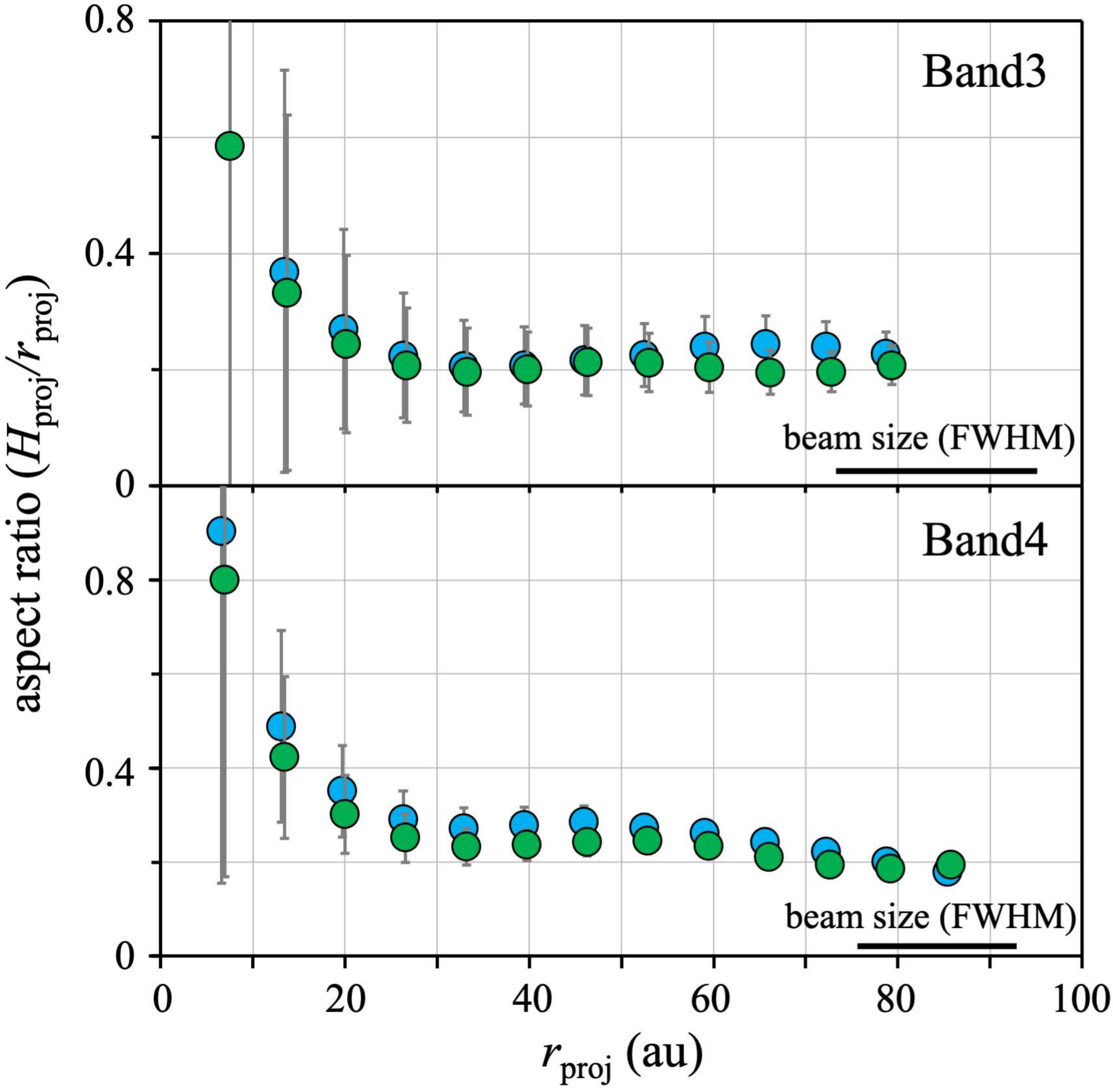}
\end{center}
\caption{Aspect ratio for disk versus radius. The error bars represent $\pm1\sigma$, which was calculated using the beam size and dust-scale height.
The green and blue circles represent the north and south directions, respectively.
}
\label{aspect}
\end{figure}

To shadow the outer disk ($r_{\rm proj}\gtrsim20$ au), the aspect ratio needs to decrease with a radius. Figure \ref{aspect} shows the observed aspect ratio for the disk as a function of the radius.
In the outer disk ($r_{\rm proj}\gtrsim30$ au), the aspect ratio seems to be constant ($\sim0.2$).
This is consistent with the increasing trend of the dust-scale height ($H\propto r^{1.0}$).
In contrast, the aspect ratio shows a decreasing trend in the inner region ($r_{\rm proj}\lesssim30$ au).
Even though the spatial resolution may be insufficient to precisely measure the aspect ratio in the inner 10-au region, the positions of the VLA clumps  ($r_{\rm proj}\sim10-20$ au) have larger aspect ratios than in the outer region. 
This is confirmed by the discovery of a substructure in the vertical direction, as discussed in Section \ref{obs:sub} (e.g., Figures \ref{fig2} and \ref{fig2dash}).
Therefore, the shadowing by the VLA clumps is a possible mechanism.

\subsection{Grain Size in Protostellar Disk}\label{dis:grain}

It is important to investigate the grain size in the disks to reveal the first step of planet formation.
One way of measuring the dust grain size is to derive the frequency dependence of the thermal dust continuum emission, because larger grains efficiently emit thermal radiation at a wavelength similar to their size \citep[e.g.,][]{dra06,ric10,tes14,kat14,bir18}.
The dust opacity index, $\beta$, provides information on grain size.

The dust opacity index was calculated as
\begin{equation}
\beta=\frac{\ln \tau_1 - \ln  \tau_2}{\ln \nu_1 - \ln \nu_2},
\label{eq:beta}
\end{equation}
where $\tau$ and $\nu$ are respectively the optical depth and frequency at each band.
To obtain the dust opacity index, the optical depth need to be measured at multiple wavelengths.
Here, we used the emission of Band 3 and Q-Band to derive the optical depth because the emission at Bands 4 and 7 was thought to be highly optically thick  at $r_{\rm proj} \lesssim50$ au.
We assume that the disk temperature is the sum of the brightness temperature for Band 4 and the background temperature of 2.7 K ($T_{\rm B}+T_{\rm bg}$).
Data for all bands were smoothed to a $\sim30$-au resolution to detect the Q-Band emission.
The spatial resolution is limited by the natural weighting of the VLA Q-Band data.

\begin{figure}[htbp]
\begin{center}
\includegraphics[width=8.cm,bb=0 0 1310 1634]{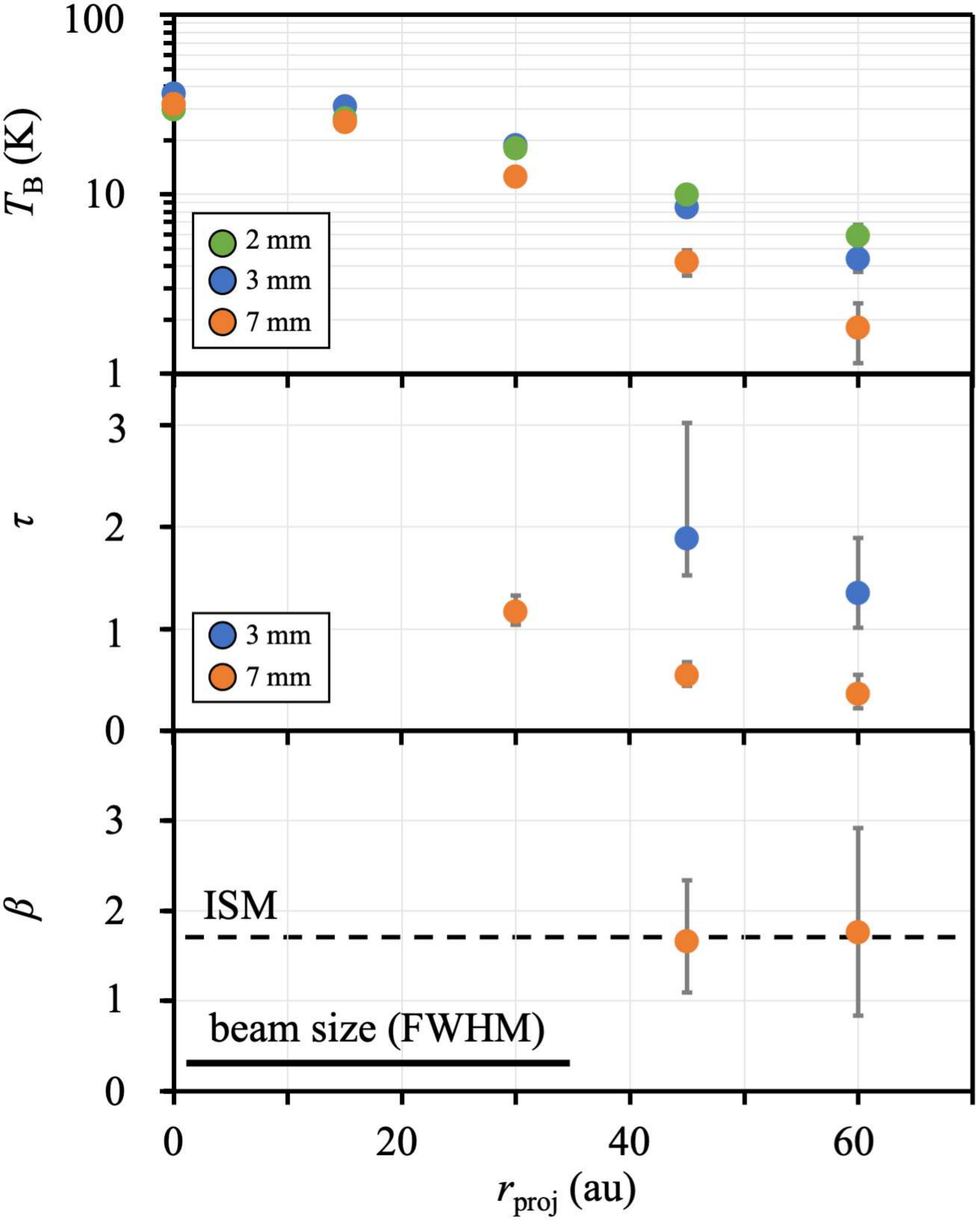}
\end{center}
\caption{Upper panel shows brightness temperatures for Band 4 and 3 and Q-Band emission versus radius. Middle panel shows radial plots of optical depths for Band 3 and Q-Band emission derived by assuming that dust temperature corresponds to Band 4. Lower panel shows dust opacity index, $\beta$, derived from optical depths for Band 3 and Q-Band emission.
The error bars represent $\pm1\sigma$, which was determined using the 5\% flux error and the rms values of the noise levels.
}
\label{beta}
\end{figure}

Figure \ref{beta} shows radial plots of the brightness temperature, optical depth, and dust opacity index.
In the central region ($r_{\rm proj}\lesssim30$ au), the optical depth for the Band 3 emission cannot be derived because the Band 3 temperature is higher than that of Band 4. 
The optical depths for Band 3 and Q-Band in the outer region were derived by using Equation (\ref{eq:radipro}), then the dust opacity indices were derived by using Equation (\ref{eq:beta}).
The derived Band 3 emission is  moderately optically thick ($\tau_{\rm 3mm}\sim1-2$), which is consistent with the low spectral index ($\alpha_{\rm 3mm-2mm}\sim2.0$).
We found that the Q-Band emission is also  optically thick  ($\tau_{\rm 7mm}\sim1.2$) around $r\sim30$ au, and become optically thin ($\tau_{\rm 7mm}\sim0.4-0.6$) at $r_{\rm proj}\gtrsim40$ au.

The derived opacity index based on these optical depths is $\beta\sim1.67^{+1.2}_{-0.7}$.
Even though the uncertainty is large, the dust opacity index, $\beta\sim1.67$, is the same as that for the ISM, which constrains the grain size to $a_{\rm max}\lesssim100$ $\mu$m by assuming that the dust grains are simple spherical particles without porosity \citep{bir18}.
Here, we note that grain size estimation strongly depends on dust models. For example, dust grains with porosity keeps $\beta\sim1.67$ up to $a_{\rm max}\lesssim1$ cm \citep{kat14,bir18}.
However, it is difficult to proceed the grain growth in such a short time scale in the L1527 protostellar disk because no substructure induced by dust accumulation is found in the outer part of the disk, $r_{\rm proj}\gtrsim40$ au.
Therefore, we suggest that significant grain growth has not yet begun in the outer region ($r_{\rm proj}\gtrsim50$ au) of the disk.

Recent coagulation simulations show that the dust grains become larger from the inner region to the outer region of a disk because the growth timescale is roughly proportional to the orbital period \citep{oha21,kob21}.
\citet{oha21} showed that the critical radius ($R_{\rm c}$) where grain growth proceeds can be calculated as
\begin{equation}
R_{\rm c}=22\left(\frac{M_\star}{0.45\ M_\odot}\right)^{1/3}\left(\frac{\zeta_{\rm d}}{0.01}\right)^{2/3}\left(\frac{t_{\rm disk}}{0.037\ {\rm Myr}}\right)^{2/3}\ {\rm au},
\label{eq:front}
\end{equation}
where the protostellar mass $M_\star$, gas-to-dust mass ratio $\zeta_{\rm d}$, and disk age $t_{\rm disk}$ are roughly the same values as those for L1527.
This means that grain growth does not occur in the outer region ($r\gtrsim22$ au), which is consistent with the large dust opacity index found around 50 au.

\subsection{Gravitational Instability and Substructure Formation}\label{dis:grav}

The origin of the VLA clumps was discussed in \citet{nak20,oha21}.
Due to the edge-on disk geometry, the substructures can be explained either by ring or spiral structures.
\citet{nak20} discussed the snow line ring because the disk temperature of $T\sim60$ K at the VLA clumps coincides with the CO$_2$ snow line, while \citet{oha21} suggested that the radius of $r_{\rm proj}\sim22$ au at the VLA clumps coincides with the growth front ring.
However, neither the snow line nor the growth front mechanism explain the shadowed region outside the ring.
To shadow the outer region, the density needs to be enhanced in the VLA clumps.

The local enhancement in density seems to be consistent with the material spiral arms formed via gravitational instability \citep{tom17}. 
Such spiral arms form in marginally unstable disks, where Toomre's {\it Q} parameter \citep{too64} is as low as 2.

The {\it Q} parameter is defined as
\begin{equation}
Q=\frac{c_{\rm s}\Omega_{\rm K}}{\pi G\Sigma},
\label{q}
\end{equation}
where $c_{\rm s}$ is the sound speed, $\Omega_{\rm K}$ is the Keplerian frequency, and $\Sigma$ is the surface density of the disk.
To measure the {\it Q} parameter, the surface density, $\Sigma$, needs to be obtained. It can be constrained by our observations.
The dust surface density, $\Sigma_{\rm dust}$, was calculated using
\begin{equation}
\Sigma_{\rm dust}=\frac{\tau_{\rm 7mm}}{\kappa_{\rm 7mm}},
\label{surface_density}
\end{equation}
where $\tau_{\rm 7mm}$ is the optical depth of the 7-mm dust continuum emission and $\kappa_{\rm 7mm}$ is the absorption opacity at a wavelength of 7 mm.
Equation (\ref{surface_density}) indicates that the optical depth and the absorption opacity are needed for the surface density.

\begin{figure}[htbp]
\begin{center}
\includegraphics[width=8.cm,bb=0 0 1776 1408]{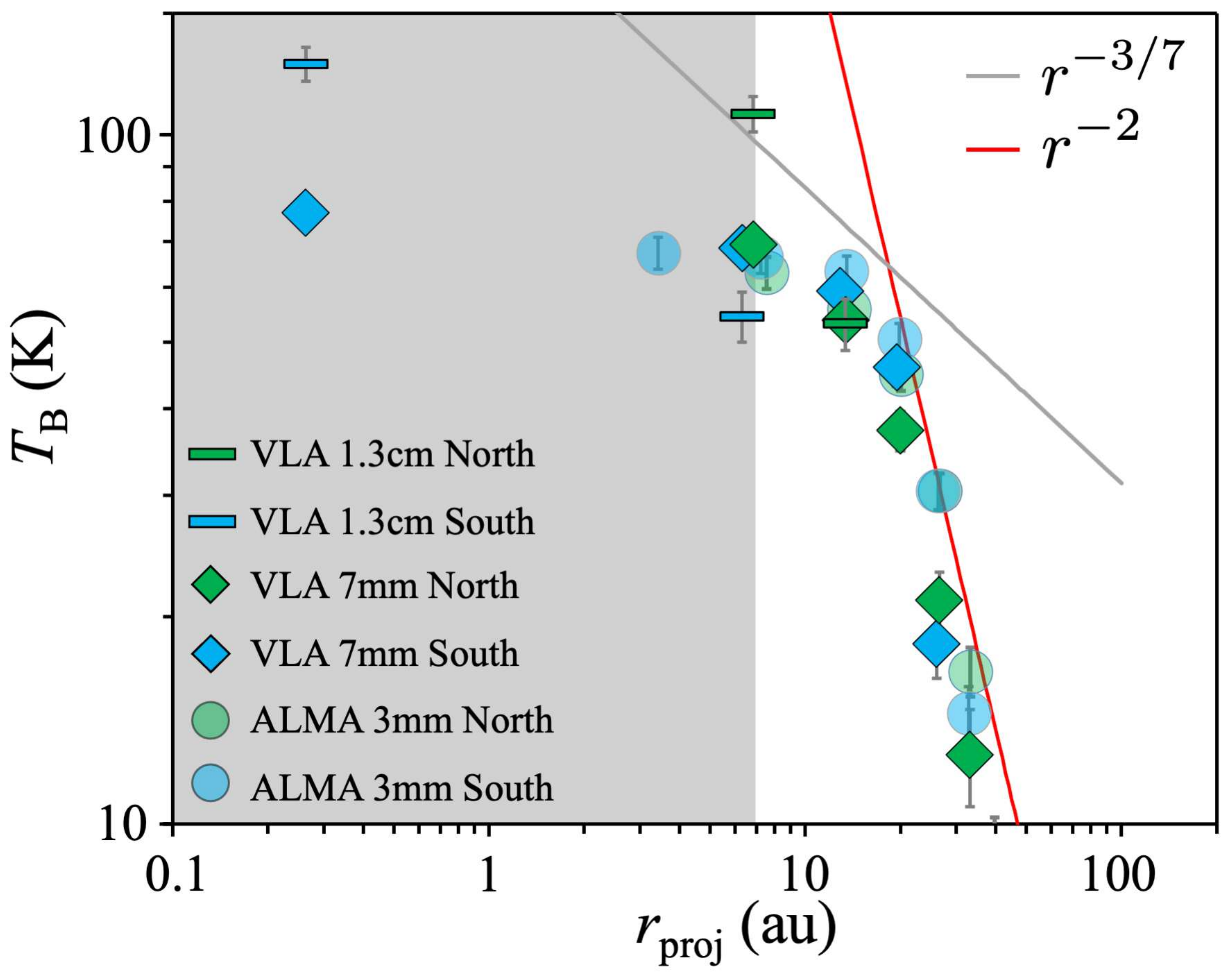}
\end{center}
\caption{Radial plots of brightness temperature of Band 3, Q-Band, and K-Band emission. The brightness temperature was derived using the higher-spatial-resolution images with a robust parameter of $-2$ shown in Figure \ref{fig2}, and smoothing to $0\farcs1$ spatial resolution. The error bars represent $\pm1\sigma$, which was calculated using the 5\% flux error and the rms values of the noise levels.
}
\label{vla_b3_v2}
\end{figure}

First, we estimate the optical depth for the Q-Band emission by assuming that the disk temperature corresponds to the brightness temperature of the Band 3 emission.
This assumption is reasonable for the VLA clump regions because the Band 3 and 4 emission is optically thick  and show similar temperatures that follow the irradiation disk model.
Figure \ref{vla_b3_v2} shows radial plots of the Band 3 and Q-Band emission by smoothing the beam size of $0\farcs1$.
Because the Q-Band emission has contaminations of the dust thermal emission and free-free emission,  Figure \ref{vla_b3_v2}  also plots the K-Band 1.3 cm emission to investigate the free-free contamination. 
The K-Band data are taken from \citet{nak20}.
Then, the  subtraction of the free-free emission in the Q-Band data is calculated as
\begin{equation}
I_{\rm 7mm,dust}=I_{\rm 7mm} - I_{\rm 1.3 cm}\Big(\frac{\nu_{\rm 7mm}}{\nu_{\rm 1.3cm}}\Big)^{-0.1},
\end{equation}
where the adopted spectral index, $-0.1$, is typical for optically thin free-free emission \citep[e.g.,][]{ang18} and is the case for L1527  \citep{nak20}.
The optical depth of the Q-Band dust thermal emission around the VLA clumps is derived to be $\tau_{\rm 7mm} \sim1.5$ after the subtraction of the  free-free emission.
Even though the temperature should be measured more precisely in future observations, the similar brightness temperatures for Band 3 and Q-Band suggest that the 7 mm emission is optically thick  at the clumps ($\tau_{\rm 7mm} >1$). 

\begin{figure}[htbp]
\begin{center}
\includegraphics[width=8.cm,bb=0 0 1607 1553]{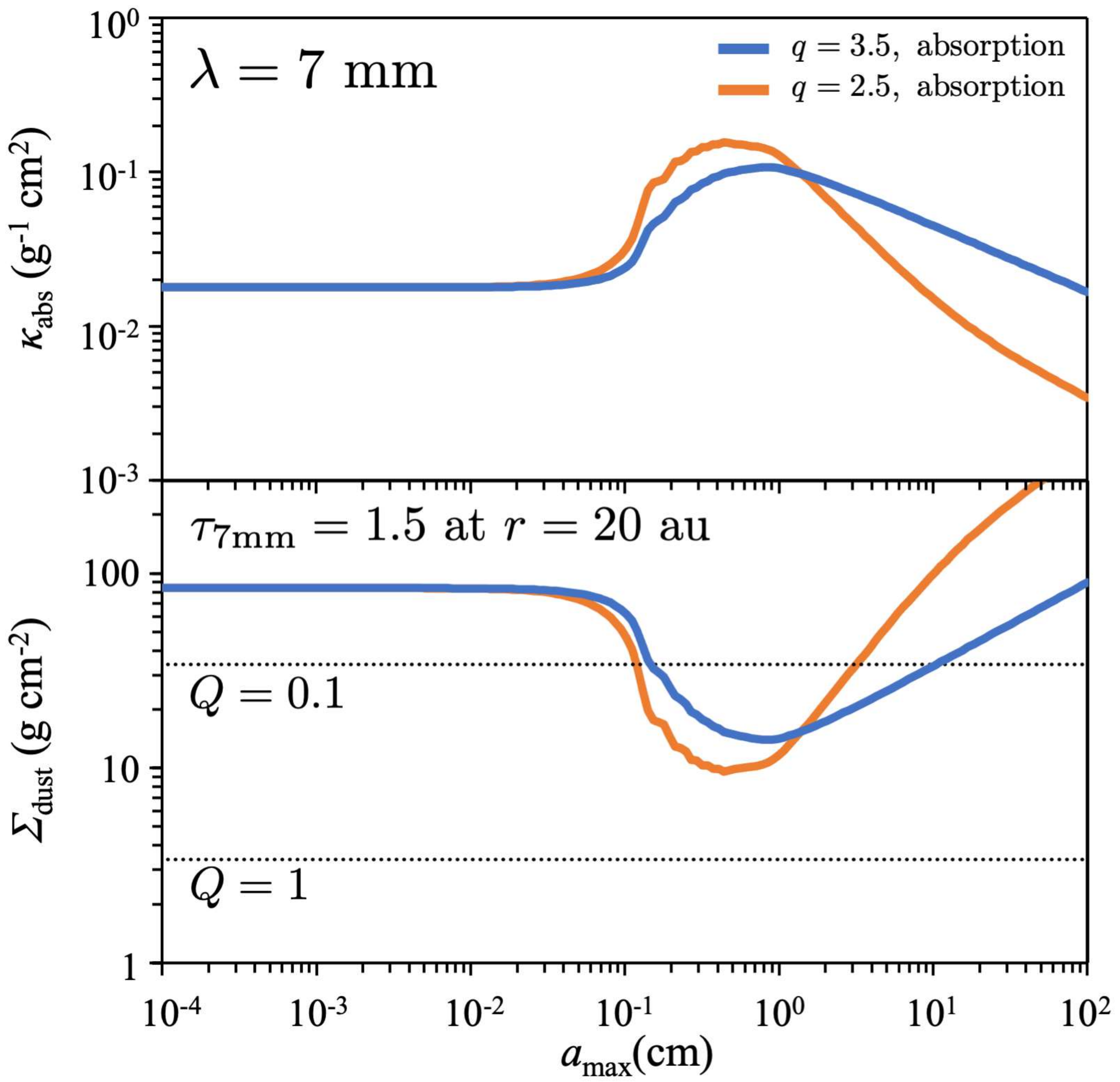}
\end{center}
\caption{Upper panel shows dust absorption opacity models for dust size distribution for $q=3.5$ and $q=2.5$ at observation wavelength of 7 mm \citep{bir18}.
Lower panel shows possible dust surface density depending on dust absorption opacity in VLA clumps. The dotted lines represent the dust surface densities where $Q=0.1$ and $Q=1$, respectively. The {\it Q} value is $Q\lesssim0.25$ for $q=3.5$.
}
\label{qpara}
\end{figure}

Second, we consider the absorption opacity, $\kappa_{\rm 7mm}$, which depends on the grain size and dust components \citep[e.g.,][]{dra06}.
We found that the dust opacity index, $\beta$, is the same as that for the ISMs in the outer region.
However, the grain size in the VLA clumps may grow due to local enhancement of the density.
Because the grain size could not be determined, we applied the upper limit of the standard dust opacity model given in \citet{bir18}, which constrains the lower limit of the dust surface density.
Figure \ref{qpara} (a) shows the dust absorption opacity as a function of grain size at an observation wavelength of 7 mm \citep{bir18}.
As reproduced in the top panel of Figure \ref{qpara}, the model given by \citet{bir18} predicts the maximum value $\kappa_{\rm 7mm}\sim0.11$ g$^{-1}$ cm$^2$ for a dust size distribution of $q=3.5$ and $\kappa_{\rm 7mm}\sim0.16$ g$^{-1}$ cm$^2$ for that of $q=2.5$. 

\begin{figure*}[htbp]
\begin{center}
\includegraphics[width=16.cm,bb=0 0 2854 1446]{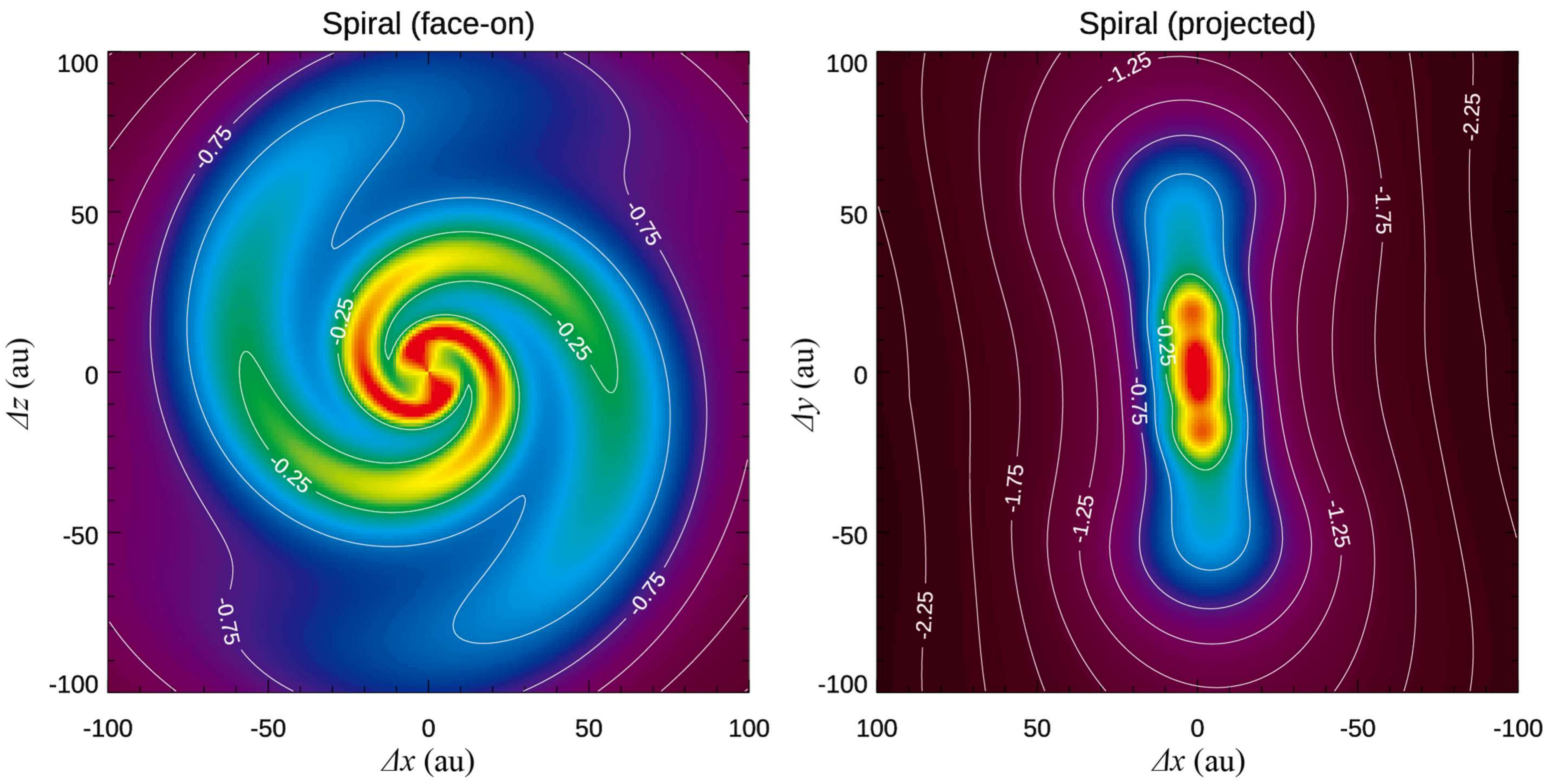}
\end{center}
\caption{The left panel shows our spiral arm model by the surface density when
seen face-on.  The right panel shows the column density along the line of sight
for the same model.  Both the inclination and position angle are assumed to
be $5 ^\circ $ in the right panel.
The contours denote $ \log _{10} \Sigma (x, y) / \max \Sigma $ and
$ \log _{10} N (- \Delta \alpha, \Delta \delta ) / \max N $ in the left
and right panels, respectively.
}
\label{spiral}
\end{figure*}

Finally, we estimate the surface density and discuss Toomre's {\it Q} parameter.
In the above discussion, the optical depth is constrained to be $\tau_{\rm 7mm}\sim1.5$  in the VLA clump regions.
In addition, the absorption opacity is assumed to be governed by the model shown in Figure \ref{qpara} (a).
Based on these parameters and Equation (\ref{surface_density}), we estimated the dust surface density as shown in  Figure \ref{qpara} (b), which constrains the lower limit of the surface density to $\Sigma_{\rm dust}\gtrsim14$ g cm$^{-2}$ for a dust grain size distribution of $q=3.5$. 
Even if we assume a dust grain size distribution of $q=2.5$, the lower limit is $\Sigma_{\rm dust}\gtrsim9.6$ g cm$^{-2}$.
By using the above dust surface densities and assuming a dust-to-gas mass ratio of 0.01, the  {\it Q} value can be calculated using Equation (\ref{q}).
The {\it Q} value was found to be $ {\it Q} \lesssim0.25$ for $q=3.5$ and ${\it Q} \lesssim0.36$ for $q=2.5$.
We show surface densities where the {\it Q} values are $Q=0.1$ and $Q=1$, respectively, in Figure \ref{qpara} (b).
The upper limit of $ {\it Q} \lesssim0.25$ indicates that the disk is gravitationally unstable at the VLA clumps.

It should be noted that the surface density used to estimate the {\it Q} value is that for a face-on disk.
In contrast, the surface density used here to estimate the {\it Q} value is that for an edge-on disk.
It is difficult to estimate the surface density for this disk when observed face-on. However, if the clumps have a spherical structure, the surface density will be mostly constant.
Even if the clumps are ring structures, the difference in the surface density observed edge-on and face-on would be similar to the aspect ratio ($H/r \sim 0.25-0.35$).
In this case, the {\it Q} value is $ {\it Q} \lesssim1.0$ for $q=3.5$.
Therefore, we conclude that the disk is gravitationally unstable in the VLA clump locations.

The gravitationally unstable condition suggests that the VLA clumps are formed by gravitational instability, such as that caused by spiral arms and fragmentation.
Spiral arms are commonly observed in \deleted{a} gravitationally unstable disks \citep{ric03,kra16}.
We investigated whether spiral arms can be observed as clumps in an edge-on disk as similar to the VLA clumps in the L1527 disk.
We applied the toy model of spiral arms by \citet{sak19} in which a geometrically thick disk contains two loosely wound spiral arms.
The assumed density distribution is given in Appendix \ref{appen2}.
The left panel of Figure \ref{spiral} displays the spiral pattern by
the surface density obtained by integrating the density along the disk
vertical direction.  The right panel denotes the column density along the
line of sight when we observe the model disk at inclination of $ 5 ^\circ$
and the position angle of $ 5 ^\circ$.  Both the surface density and column density
are normalized by their maximum values.  The column density has three clumps
along the disk as in the brightness distribution at the VLA Q-Band.
Thus, the spiral arm model is applicable to the inner part of the disk.

Fragmentation may also occur in the spiral arms and produce several clumps if the disk is sufficiently massive.
\citet{tak16} investigated the conditions required for fragmentation in a disk using two-dimensional numerical simulations.
They suggested that Toomre's {\it Q} parameter in the spiral arms is essential for fragmentation. The spiral arms fragment when $Q < 0.6$ in the spiral arms. 
If the VLA clumps are fragments in spiral arms rather than a ring structure, the derived {\it Q} value is ${\it Q}\lesssim0.25$, consistent with the fragment threshold of $Q < 0.6$. These results suggest that the substructure observed in the VLA can be the clumps formed by gravitational instability.

If the VLA substructures (Clump-N and S) are real clumps, the estimation of these masses is important for planet formation or disk evolution.
We estimated the dust mass of the VLA clumps (Clump-N and S) using 
\begin{equation}
M_{\rm dust}=\frac{F_{\rm 7mm}}{\kappa_{\rm 7mm} B_\nu (T_{\rm dust})},
\label{mass}
\end{equation}
where $F_{\rm 7mm}$ is the flux density, $\kappa_{\rm 7mm}$ is the dust opacity, and $B_\nu$ is the Planck blackbody function.
We used the maximum value $\kappa_{\rm 7mm}\sim0.11$ g$^{-1}$ cm$^2$ for the dust opacity to constrain the lower limit of the dust mass.
The derived dust mass is $M_{\rm dust}\gtrsim0.12$ $M_{\rm J}$ and $M_{\rm dust}\gtrsim0.17$ $M_{\rm J}$ for the northern and southern clumps, respectively.
Here, the thermal dust component at 7 mm is used by subtracting the free-free emission component \citep{nak20}.

The existence of planets in Class II protoplanetary disks was suggested by a kinematic analysis of molecular  lines \citep{pin18,tea18}. 
In addition, a circum-planetary disk has been identified in the protoplanetary disk around PDS 70 \citep{kep18}.
These results indicate that planet formation starts in Class I disks or at a much earlier stage.
Indeed, recent ALMA observations of Class I protostellar disks indicated that planet formation may already occur in the Class I disk-forming stage \citep{seg20,alv20}.
Our results may support such early planet formation in Class 0/I protostellar disks. The L1527 disk can be massive enough to be enhance the gravitationaly instability, which might induce planet formation in future \citep[e.g.,][]{gam01}.
\citet{xu22} also suggested that the majority of disks are likely to be gravitationally unstable.

Rather than the planet formation, it may also be possible for the VLA clumps to accrete to the central star and cause an accretion burst such as in the case of FU Orionis, which is a low-mass young stellar object that shows an accretion outburst with an accretion rate that increases from $\dot{M_\star}\sim(10^{-7}-10^{-8})$ $M_{\odot}$ yr$^{-1}$ to $(10^{-5}-10^{-4})$ $M_{\odot}$ yr$^{-1}$ over several decades to 100 years \citep{aud14}.
The masses of the VLA clumps are sufficient to increase the accretion rate if these clumps accrete to the central star via migration \citep[e.g.,][]{lin86,war97,tan02,vor15}.
Thus, the VLA clumps might be the origin of future accretion bursts.

As summarize the discussion of this subsection, the L1527 disk is found to be gravitationally unstable ($Q\lesssim1.0$). Therefore, the VLA clumps would be formed by the disk gravitational instability.
These VLA clumps may be the origin of gas-giant planets or accretion burst by accreting onto the central star.

\subsection{Caveat and Other Possibilities for Substructure Formation}

In the above discussion, we showed that the disk is gravitationally unstable because the {\it Q} value is ${\it Q}\lesssim1.0$.
However, the disk surface density for estimating the {\it Q} value strongly depends on the dust opacity ($\kappa_{\rm abs}$).
In this study, we used the standard opacity model developed by \citet{bir18}.
There are some opacity models that have larger dust opacities, such as that in \citet{jag98,dra03,zub96} and some carbonaceous dust models in \citet{bir18}.
If we apply these opacity models, it is possible for the disk to be gravitationally stable with ${\it Q}>2.0$ only when the dust grains are larger.
Although we suggested that gravitational instability is the likely origin of the substructure, other mechanisms may have created the shadowing effect at $r\sim20$ au.
By taking into account the shadowing by the VLA clumps, a dust pile-up mechanism is needed to form the disk substructures such as the dead zone \citep{flo15}.
Further observations with higher spatial resolution and sensitivity are needed to confirm the origin of the substructures.

\section{Summary}\label{sec:sum}

We have analyzed multi-wavelength dust continuum emission toward the protostellar disk around the Class 0/I protostar L1527 IRS.
The dust continuum data cover a wide wavelength range of 0.87, 2.1, 3.3, and 6.8 mm obtained by the ALMA and VLA observations, and have angular resolutions of $0\farcs025-0\farcs16$ (corresponding to $3.4-20$ au). The main results are listed below.

\begin{itemize}

\item The protostellar disk was shown to be an edge-on structure aligned in the north-south direction based on the ALMA observations. This structure is similar to that derived from previous observations \citep[e.g.,][]{aso17,sak17}.
In contrast, the VLA observations show that the 6.8-mm emission is only detected within 30 au, which is attributed to the sensitivity limit.

\item The peak temperature increases with increasing wavelength, which is explained by the optical depth effect; emission at shorter wavelengths is  optically thicker and is obscured by the outer cold annuli. In contrast, the higher brightness temperature found for the Band 7 emission in the outer part of the disk ($r_{\rm proj}\gtrsim50$ au) suggests the existence of an envelope component or accretion shock to the edge of the infant disk.

\item Based on an analysis of the optical depths associated with the Band 3 and Q-Band emission, the dust opacity index, $\beta$, was determined to be $\beta\sim1.7$ at $r_{\rm proj}\sim50$ au, suggesting that significant dust grain growth has not yet begun. This is consistent with the growth front model developed by \citet{oha21} because the growth front model predicts that the critical radius ($R_c$) where the grain growth proceeds can be calculated as 
\begin{equation}
R_{\rm c}=22\left(\frac{M_\star}{0.45\ M_\odot}\right)^{1/3}\left(\frac{\zeta_{\rm d}}{0.01}\right)^{2/3}\left(\frac{t_{\rm disk}}{0.037\ {\rm Myr}}\right)^{2/3}\ {\rm au},
\label{eq:front}
\end{equation}
where the protostellar mass $M_\star$, gas-to-dust mass ratio $\zeta_{\rm d}$, and disk age $t_{\rm disk}$ are roughly the same values as those for L1527.
This means that grain growth does not occur in the outer region ($r\gtrsim22$ au).

\item The disk temperature seems to show a power-law profile with an index of $T\propto r_{\rm proj}^{-3/7}$ inside a 20-au radius, and a steeper power-law profile with an index of $T\propto r_{\rm proj}^{-2}$ at $r_{\rm proj}\gtrsim20$ au.
The inner temperature can be explained by an irradiation model, while the steep drop in temperature toward the outer region is caused by a shadowing effect by the VLA clumps. 

\item  The VLA clumps form via gravitational instability because the {\it Q} value is estimated to be $Q\lesssim1.0$.
The derived dust mass for the VLA clumps is $\gtrsim0.1$ $M_{\rm J}$.
Thus, we suggest that Class 0/I disks can be sufficiently massive to be gravitationally unstable, which might be the origin of gas-giant planets in a 20-au radius. 

\end{itemize}

We gratefully appreciate the comments from the anonymous referee that significantly improved this article.
This paper makes use of the following ALMA data: ADS/JAO.ALMA\#2019.1.01695.S, \\
ADS/JAO.ALMA\#2017.1.00509.S,\\ ADS/JAO.ALMA\#2016.A.00011.S. ALMA is a partnership of ESO (representing its member states), NSF (USA) and NINS (Japan), together with NRC (Canada), MOST and ASIAA (Taiwan), and KASI (Republic of Korea), in cooperation with the Republic of Chile. The Joint ALMA Observatory is operated by ESO, AUI/NRAO and NAOJ. The National Radio Astronomy Observatory is a facility of the National Science Foundation operated under cooperative agreement by Associated Universities, Inc.

This project is supported by pioneering project in RIKEN (Evolution of Matter in the Universe) and a Grant-in-Aid from Japan Society for the Promotion of Science (KAKENHI: Nos. 22H00179, 22H01278, 21K03642, 20H04612, 20H00182, 20K14533, 18H05436, 18H05438, 17H01103, 17H01105).

H.B.L. is supported by the Ministry of Science and
Technology (MoST) of Taiwan (Grant Nos. 108-2112-M-001-002-MY3 and 110-2112-M-001-069-).

Data analysis was in part carried out on common use data analysis computer system at the Astronomy Data Center, ADC, of the National Astronomical Observatory of Japan.

\facilities{ALMA, JVLA}

\software{CASA \citep[][]{mcm07}, RADMC-3D \citep{dul12}
          }




\appendix

\section{Simulation observations for the disk temperature}\label{appen}

In Section \ref{dis:temp}, we discuss the steep temperature of $T\propto r^{-2}$ due to the shadowing.
However, one may consider that the steep temperatures are caused by the resolved-out effect of the ALMA extended configurations.

To investigate the resolved-out issue, we conduct the simulation observations in CASA {\it simobserve} at ALMA Band 4.
The disk model is created by RADMC-3D \citep{dul12}.
The temperature and dust surface density are assumed to be $T=224 (r/{\rm AU})^{-3/7}$ and $\Sigma_{\rm dust}=10$ g cm$^{-2}$, respectively, to mimic the observed optically thick emission.
The disk size is assumed to be $r_{\rm out}=100$ au.
Note that the temperature profile assumes the irradiation model, same as Figures \ref{temp} and \ref{temp2}, but the dust surface density  should be made with any optically thick model.
We simply assume a constant dust surface density independent of radius.
The dust size is simply assumed to be $a_{\rm max}=0.1$ $\mu$m because the grain growth does not proceed yet discussed in Section \ref{dis:grain}.
The inclination is set to be $15^\circ$ to mimic the large dust scale height.

The CASA {\it simobserve} is applied to the model image with a total integration time of 120 min with the most extended configuration in Cycle 8 (alma.cycle8.10.cfg), which is the same observing setup with our observations.
A Precipitable Water Vapor (PWV) of 0.5 mm is used to input the noise. 
After the measurements sets (MS file) are created by {\it simobserve}, we map the simulation image in CASA {\it tclean}.

\begin{figure*}[htbp]
\begin{center}
\includegraphics[width=16.cm,bb=0 0 2065 1687]{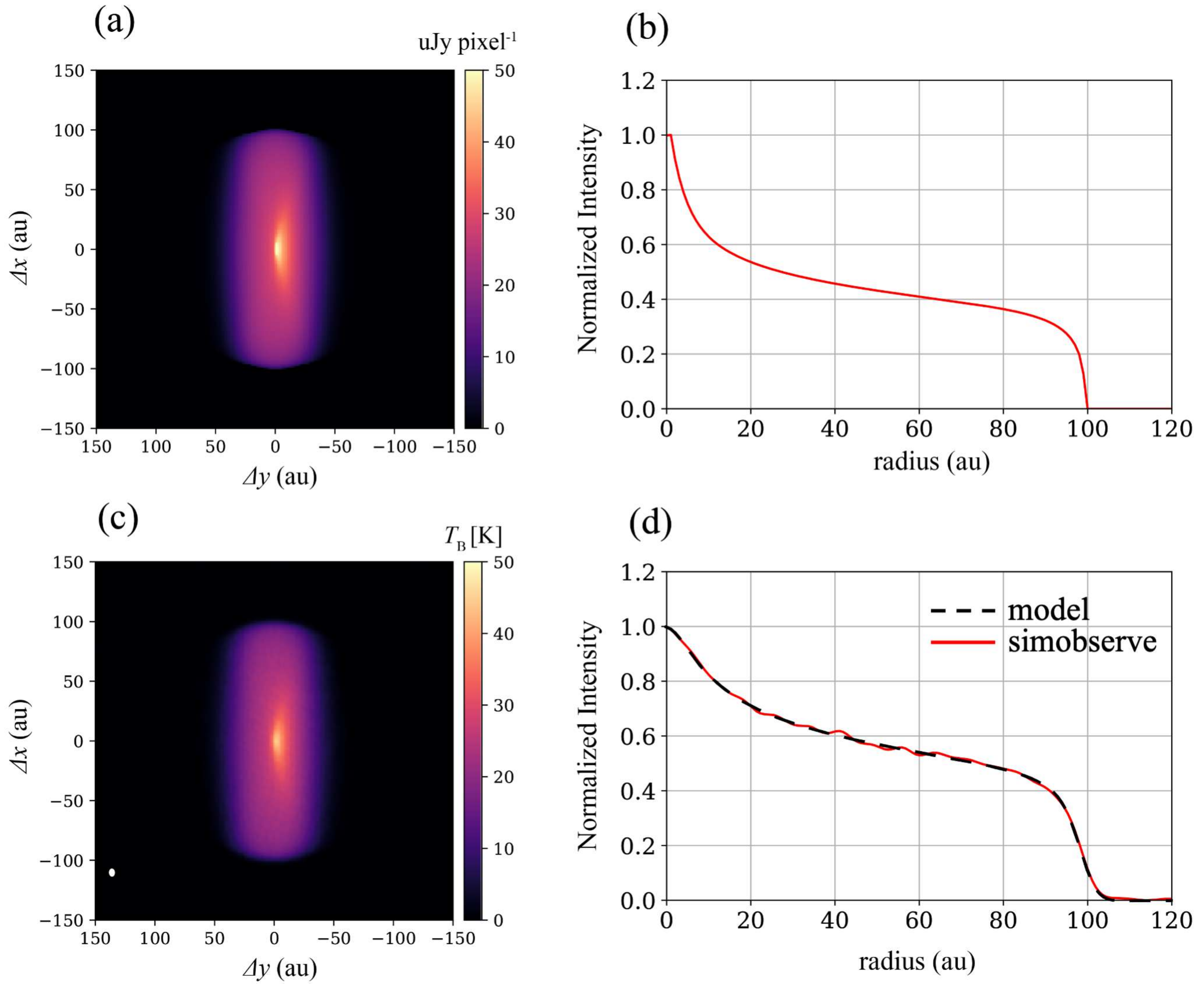}
\end{center}
\caption{Disk image (a) and radial plot of normalized intensity (b) for the model calculation are shown, respectively.
The same image and radial plot are also shown in (c) and (d) but for the simulation observations.
The panel (d) also plots the intensity distribution of panel (b) smoothed by the beam size in the black dashed lines for comparison.
}
\label{a1}
\end{figure*}

Figure \ref{a1} (a) and (b) show the model image and the radial plot of the normalized intensity, respectively.
Furthermore, Figure \ref{a1} (c) and (d) shows the same image and radial plot but for the simulation observations.
The beam size of the simulation observations is $0\farcs049\times0\farcs036$ shown in the bottom-left corner of Figure \ref{a1} (c).
Figure \ref{a1} (d) also plots the radial distribution of the model smoothed by the beam size for comparison.
Both of the model and simulation observations show almost the same intensity distributions, although some intensity fluctuations and flatter power-law profile are found in the simulation observations due to the noise and beam convolution.
Because intensity profile does not change by the simulation observations, we conclude that the observed steep temperature profile is not caused by the resolved-out issue.

\section{Spiral Model}\label{appen2}

Our spiral model is essentially the same as that of \cite{sak19}, 
although the model parameters are slightly different.
To describe the model, we use the Cartesian coordinates, $(x, y, z) $, 
of which origin coincides with the protostar.
The $x$-axis is parallel to the disk major axis on the sky, while
the $ z $-axis is perpendicular to the disk plane. The inclination
and position angle of the disk are assumed to be $ i = 5 ^{\circ} $
and $ \mbox{P.A.} _{\rm d} = 5 ^{\circ} $, respectively.  Then
the Cartesian coordinates are related to the observation by
\begin{eqnarray}
x & = & D \left(\Delta \alpha \sin \mbox{P.A.} _{\rm d} 
+ \Delta \delta \cos \mbox{P.A.} _{\rm d} \right) , \\
y & = & D \sin i \left( \Delta \alpha \cos \mbox{P.A.}_{\rm d} - \Delta \delta 
\sin \mbox{P.A.} _{\rm d} \right) + s \cos i  , \\
z & = & D \cos i \left( \Delta \alpha \cos \mbox{P.A.} _{\rm d} 
- \Delta \delta \sin \mbox{P.A.} _{\rm d} \right) - s \sin i ,
\end{eqnarray}
where $ D $ and $ s $ denote the distance to L1527 and coordinate
along the line of sight, respectively. For later convenience, we
define the spherical coordinates,
\begin{eqnarray}
r & = & \sqrt{x ^2 + y ^2 + z ^2} , \\
\theta & = & \cos ^{-1} \frac{z}{r} , \\
\varphi & = & \tan ^{-1} \frac{y}{x} .
\end{eqnarray}

Using the coordinates defined above, we express the
density distribution as follows.
\begin{eqnarray}
\rho(x, y, z) & = & \rho _0 (r, \theta) 
\exp \{a(r) \cos \left\{  2 \left[ \varphi -  \phi (r) \right] \right\} \\
\rho _0 (r,z) & = & A \exp \left\{ - \sqrt{\displaystyle \left(
\frac{r}{25~\mbox{au}} \right) ^2 +1} - \frac{1}{2} \left[ \frac{r \cos \theta}{H(r)} \right] ^2 \right\} ,\\
H (r) & = &  6.584 \left[ \left( \frac{r}{40~\mbox{au}} \right) ^4 +1 \right] ^{0.4}
\mbox{au} ,\\
\phi _0 (r) & = & p \ln \left[ \frac{r}{10~\mbox{au}} , 1 \right] - \chi , \\
a (r) & = & 0.4 \min \left[ \left( \frac{r}{10~\mbox{au}} \right) ^2, 1 \right] .
\end{eqnarray}
Here, the parameters, $ A $, $ p $, and $\chi $ denote the density at the origin, pitch angle and spiral phase, respectively. Since we are interested only
in the relative density variation, the value of $ A $ is not yet fixed.
The rest parameters are chosen to be $ p = 3.0 $ and $ \chi = 103^\circ $.

The left panel of Figure \ref{spiral} denotes the surface density of the
disk,
\begin{eqnarray}
\Sigma (x, y) & = & \int \rho (x, y, z) dz .
\end{eqnarray}
The right panel denotes the column density along the line of sight,
\begin{eqnarray}
N (\Delta \alpha, \Delta \delta ) & = & \int \rho (x, y, z) ds ,
\end{eqnarray}
which mimics the Q-band image.  
The surface density is normalized by the maximum value on each panel.
The contours denote $ \log _{10} \Sigma (x, y) / \max \Sigma $ and
$ \log _{10} N (- \Delta \alpha, \Delta \delta ) / \max N $ in the left
and right panels, respectively.

\end{document}